\documentclass[aps,pra,showpacs,superscriptaddress]{revtex4-2}
\usepackage{xcolor}
\usepackage{graphicx}
\usepackage{epstopdf}
\usepackage{placeins}
\usepackage{natbib}
\usepackage{amsmath}
\usepackage[percent]{overpic}
\usepackage{hyperref}
\DeclareMathOperator{\csch}{csch}

\hypersetup{
	colorlinks=true,linkcolor=blue,citecolor=blue,
	filecolor=blue,urlcolor=blue,breaklinks=true
}

\RequirePackage{color}

\begin{document}

\title{Impact of the Sagnac Effect on Thermodynamic and Magnetocaloric Properties of a Rotating Two-Dimensional Electron Gas}

\author{Cleverson Filgueiras}
\author{Moises Rojas}
\author{Vinicius T. Pieve}
\affiliation{
 \quad Departamento de F\'{\i}sica (DFI),
Universidade Federal de Lavras (UFLA), Caixa Postal 3037,
37200--000, Lavras, Minas Gerais, Brazil; \\}
\author{Edilberto O. Silva}
\affiliation{Departamento de F\'{\i}sica, Universidade Federal do Maranh\~{a}o, 65085-580 S\~{a}o Lu\'{\i}s, MA, Brazil}
\date{\today}

\begin{abstract}
This work investigates the impact of the Sagnac effect on the thermodynamic properties of a non-interacting two-dimensional electron gas (2DEG) in a rotating sample under the influence of a uniform magnetic field. We derive an analytical expression for the energy spectrum using an effective Hamiltonian that incorporates inertial forces, and then apply canonical ensemble statistical mechanics to evaluate thermodynamic quantities. The results show that rotation modifies the energy levels, the application of a magnetic field leads to the formation of Landau levels further altered by rotation and gravitational mass, and thermodynamic quantities (internal energy, specific heat, free energy, entropy, magnetization, and magnetocaloric effect) exhibit a strong dependence on these parameters. In particular, the difference between effective mass $m^*$ and gravitational mass $m_G$ influences magnetization and the magnetocaloric effect, with negative rotations potentially inducing a cooling effect when these masses are distinct. We conclude that rotational effects and effective mass properties are crucial for understanding the thermodynamics of electronic systems under magnetic fields, with implications for thermal modulation in semiconductor materials.

\end{abstract}
\maketitle

\section{Introduction}

The Sagnac Effect \cite{sagnac1913ether,SagnacEffect,SagnacRevModPhys.39.475} is a fundamental physical phenomenon with profound implications for both theoretical physics \cite{SagnacTheory1,SagnacDiracFermions,SagnacRelativity} and technological applications \cite{SagnacGravityLaser,Culshaw_2006}. Its sensitivity to rotation, gravity, and the properties of matter makes it a valuable tool for investigating crucial questions in fundamental physics, such as the nature of dark matter \cite{SagnacSolarSystemProbeDarkMatter} and gravitomagnetic effects \cite{SagnacGravitomagentism}, while also serving as the basis for high-precision rotational sensors used in various fields \cite{SagnacSensor1,SagnacSensor2}. The continuous development of Sagnac interferometers, both optical \cite{SagnacInterferometer} and matter-wave-based \cite{SagnacMatterWaves}, promises significant advancements in our understanding of the universe and sensing technologies.

The Sagnac effect is closely related to inertial forces, as it can be used to study and determine them. Inertial forces, such as the Coriolis and Euler forces, manifest in rotating or non-uniformly rotating reference frames. Inertial effects play a crucial role in quantum mechanics \cite{rot1,rot2,rot3,AmaroNeto:2024hfp}. The well-established analogy between inertial forces on massive particles and electromagnetic forces on charged particles provides an insightful perspective \cite{10.1119/1.19503}. The Coriolis force, for instance, acts on a particle with mass $ m $ in a manner analogous to the magnetic force on a charged particle.
In contrast, the centrifugal force affects the particle within a rotating reference frame. For a spinless particle, the combined action of Coriolis and centrifugal forces in the quantum Hamiltonian results in a coupling between the particle's angular momentum and the system's rotation \cite{10.1119/1.19503,fischer2001hall}. Even in the absence of the centrifugal force, the Landau levels in the system still exhibit coupling between the particle's angular momentum and rotation. Furthermore, if the rotation changes steadily over time, the Euler force must also be accounted for in the analysis.

Among the quantum systems where rotational and electromagnetic effects become particularly relevant, the two-dimensional electron gas (2DEG) stands out as a model system with rich physics and technological potential\cite{devreese2012physics}. A 2DEG is a system in which electrons are confined to move in only two spatial dimensions, typically formed at the interface of semiconductor heterostructures such as GaAs/AlGaAs, or in devices like metal-oxide-semiconductor field-effect transistors (MOSFETs)\cite{goerbig2009quantum}. The high mobility of these systems enables the observation of quantum phenomena such as the Quantum Hall Effect\cite{YEH2023149}, making them essential for the study of low-dimensional systems and strongly correlated electrons\cite{RevModPhys.75.1101}. Under strong magnetic fields, these systems exhibit remarkable quantum properties, including conductance quantization and the formation of Landau levels\cite{aoki2013physics}. They have shown significance in optoelectronic applications as well\cite{chen2025optoelectronic}. The development of semiconductor nanomaterials and the exploration of quantum effects at the nanoscale are driving significant advancements in high-performance devices such as efficient solar cells, color micro-LEDs, advanced sensors, and innovative 3D displays\cite{shim2017electronic}.

When considering electrons in materials, an intriguing question arises regarding the distinction between three types of mass: inertial, gravitational, and effective mass. Inertial mass reflects an object’s resistance to changes in its motion, while gravitational mass relates to the gravitational attraction between bodies. According to the equivalence principle in general relativity, inertial and gravitational masses are equivalent. In contrast, the effective mass, which describes an electron’s response to electromagnetic forces within a material, can deviate from both inertial and gravitational masses, depending on the material's structure \cite{WANG2024107117}. The effective mass is derived from the curvature of the material’s energy band structure and can vary significantly, largely independent of gravitational effects. Despite this variability in the effective mass due to electromagnetic interactions, the gravitational mass of electrons remains constant, regardless of the medium in which they are situated.

According to \cite{WANG2024107117}, both effective mass $ m^* $ and gravitational mass $ m_G $ should be considered within the Schrödinger equation. For electrons in semiconductors, these masses differ, which can lead to notable shifts in energy levels and, consequently, in the physical properties of the system under study \cite{FILGUEIRAS2025169858}. Although \cite{WANG2024107117} does not explicitly frame the discussion, the work conceptually addresses the distinction between these three masses, emphasizing that while gravitational mass is associated with an accelerated reference frame, effective mass governs electron behavior within materials. Notably, the equivalence principle does not directly compare gravitational and effective masses; a conceptually distinct test would be required to evaluate them.

Our aim is to investigate thermodynamic properties and the magnetocaloric effect (MCE) for a two-dimensional electron gas (2DEG) subject to inertial effects associated with the Sagnac effect. Since two models are presented in the literature, we will analyze both contexts as mentioned above. We will follow the general model, for which $ m^* \neq m_G $, obtaining the equivalent results simply by considering these masses equal.

In our understanding, although the arguments presented in Ref. \cite{WANG2024107117} are strong, we still believe that the effective theory for the case $ m^* \neq m_G $ should be investigated through first-principles calculations. However, experiments and/or simulations involving thermodynamic properties could shed light on this matter.

This work is organized as follows. In Section \ref{sec2}, we derive the energy levels for a rotating 2DEG under a uniform magnetic field applied perpendicular to the plane of rotation. In Section \ref{sec3}, we introduce the thermodynamic properties to be analyzed. Section \ref{sec4} presents the results and a detailed discussion of their implications. Finally, our conclusions are summarized in Section \ref{sec5}.

\section{Energy levels for a rotating 2DEG with a uniform magnetic field applied perpendicular to the rotating plane} \label{sec2}

Reference \cite{WANG2024107117} addresses the distinction between inertial, gravitational, and effective mass. Inertial mass, tied to an object's resistance to motion, differs from gravitational mass, which relates to gravitational attraction, as the Equivalence Principle describes. While gravitational and inertial masses are equivalent in general relativity, effective mass pertains to how electrons respond to electromagnetic forces in materials. For semiconductors, the effective mass $m^*$ is derived from the material's energy band structure, influenced by the medium rather than gravitational forces. The equation $m^* = \hbar^2 \left( d^2 E/dk^2 \right)^{-1}$ captures this dependence, where $E$ is energy, $k$ is the wave vector, and $\hbar$ is the reduced Planck constant. Despite variations in $m^*$, the gravitational mass $m_G$ remains constant, regardless of the medium. Therefore, $m_G \neq m^*$ generally, unless in cases like simple metals, where all masses coincide. Discussions of effective mass often focus on electromagnetic interactions, which do not alter gravitational mass. Further study of these concepts, particularly in the context of massless fermions and accelerated reference frames, would be insightful.

We now turn to the effective theory describing electrons under inertial forces, namely the Coriolis and centrifugal forces. The quantum Hamiltonian in cylindrical coordinates for a particle on a rotating disk with angular velocity $\boldsymbol{\Omega} = \Omega \boldsymbol{\hat{z}}$ is given by \cite{10.1119/1.19503,WANG2024107117}
\begin{equation}
H = \frac{[\mathbf{P} - m_G (\boldsymbol{\Omega} \times \mathbf{r})]^2}{2m^*} - \frac{m_G}{2} (\boldsymbol{\Omega} \times \mathbf{r})^2,\label{hamilton}
\end{equation}
where $m_G (\boldsymbol{\Omega} \times \mathbf{r})$ and $m_G (\boldsymbol{\Omega} \times \mathbf{r})^2$ account for the Coriolis and centrifugal forces, respectively. Ref. \cite{WANG2024107117} derives these rotation-induced inertial forces from the Sagnac phase, which arises from the time difference in travel paths in a rotating interferometer. The phase shift is proportional to the system's angular velocity $\Omega$ and cross-sectional area $S$
\begin{equation}
\Delta \phi = \frac{2 m_G \Omega S}{\hbar},
\end{equation}
where $m_G$ is the electron’s gravitational mass and $S$ is a cross-sectional area. According to the equivalence principle, $m_G$ refers to the gravitational mass that appears in the terms induced by the accelerated reference frame, while $m^*$ represents the effective mass, as previously discussed. The two masses are not directly compared, as the equivalence principle remains valid. A different experimental setup is needed to test the distinction, such as investigating the thermodynamic properties of a rotating 2DEG. 

In the presence of a magnetic field  $\mathbf{B}=B \boldsymbol{\hat{z}}$, Eq. (\ref{hamilton}) can be rewritten as
\begin{eqnarray}
H=\frac{[{\bf P}-q{\bf A}-m_G({\bf \Omega}\times{\bf r})]^2}{2m^*}-\frac{m_G({\bf \Omega}\times{\bf r})^2}{2}+qV(r).\label{hamiltonianageral}
\end{eqnarray}
We consider a  flat sample with the line element in polar coordinates, that is,
\begin{equation}
ds^2 = dr^2 + r^2 d\theta^2, \label{metric}
\end{equation}
where $r\geq0$ and $0\leq \theta \leq 2\pi$
For a perpendicular constant magnetic field, the scalar and vector electromagnetic potentials are, respectively, given by
\begin{align}
V(r)&=-\frac{\Omega Br^2}{2},\\
\mathbf{A}&=(0,\frac{Br}{2\color{red}},0).
\end{align}
The electric field associated with the scalar potential appears when the applied magnetic field is transformed into the rotating frame. 

From Eq. (\ref{hamiltonianageral}), the Schr\"odinger equation $H\Psi(r,\theta)=E\Psi(r,\theta)$, with $\psi\equiv R(r)e^{-i\ell\theta}$, leads to 
\begin{eqnarray}
r^2R''+rR'+(-\sigma^2r^4+\gamma^2 r^2-\ell^2)R=0, \label{eqemr}
\end{eqnarray}
where 
\begin{eqnarray}
\sigma^2=\frac{m^* \omega_c^2}{4\hbar^2}+\frac{\left(m_G-m^*\right)m_G \Omega^2}{\hbar^2}+\frac{\left(m_G-m^*\right)m^* \omega_c\Omega}{\hbar^2}
\end{eqnarray}
and
\begin{eqnarray}
\gamma^2=\frac{2m}{\hbar}\left(\frac{E}{\hbar}-\frac{\omega_c \ell}{2}-\frac{m_G \Omega \ell}{m^*}\right).
\end{eqnarray} 
Writing $\sigma r^2=\xi $ and looking at the asymptotic limit as $\xi\rightarrow\infty$, the general solution to this equation will be given in terms of  $\rm M(a,b,\xi)$, namely {\it a confluent hypergeometric function of the first kind}  \cite{abramo}, 
{\small\begin{equation}
R\equiv R_{\ell}\left( \xi \right)=a_{\ell}e^{-\frac{\xi}{2}}\xi^{\frac{|\ell|}{2}}{\rm M}\left(-\frac{\gamma^{2}}{4\sigma}+
\frac{|\ell|}{2}+\frac{1}{2} ,1+|\ell|
,\xi\right)
+b_{\ell}e^{-\frac{\xi}{2}}\xi^{-\frac{|\ell|}{2}}{\rm M}
\left(-\frac{\gamma^{2}}{4\sigma}-\frac{|\ell|}{2}+\frac{1}{2} ,1-|\ell|
,\xi\right),
\label{general_sol_2_HO}
\end{equation}}
where $a_{\ell}$ and $b_{\ell}$ are, respectively, the coefficients of the {\it regular} and {\it irregular}
solutions. Notice that the term {\it irregular} stems from the fact that it diverges as $\xi \rightarrow 0$. Then, we take $b_{\ell}\equiv0$.

To have a finite polynomial function (the hypergeometric series has to be convergent to have a physical solution), the condition ${{\rm a}=-n}$, where $n$ is a positive integer number,  has to be satisfied. From this condition, the  possible values for the energy are given by
\begin{equation}
E_{n,l}=\hbar\left(\frac{\omega_c }{2}+\frac{m_G\Omega}{m^*}\right)\ell+\hbar\sqrt{\omega_{c}^2+4m_G\frac{(m_G-m^*)}{m^{*2}}\Omega^2+4\omega_c\Omega\frac{(m_G-m^*)}{m^*}}\left(n+\frac{|\ell |}{2}+\frac{1}{2}\right),
\label{Energyspectrum}
\end{equation}
where $\omega_c\equiv qB/m^*$ is the cyclotron frequency.
The energy spectrum above is impacted by the Sagnac effect represented by the mass term $m_G$. We remark that the fields $B$ and $\Omega$ are externally tunable parameters. The presence of rotation by itself breaks the degeneracy between states with opposite angular momentum. It introduces energy shifts depending on the rotation ${\Omega}$ \cite{BRANDAO201555}, which were analyzed in Ref. \cite{BRANDAO201555,Brandao2017}, with $m_{G}\equiv m^*$. Positive rotation ($\Omega > 0$) refers to counterclockwise motion, while negative rotation ($\Omega < 0$) refers to clockwise motion, following the right-hand rule convention with the angular velocity vector pointing along the $+z$ direction. We can rewrite the energy levels in Eq. (\ref{Energyspectrum}) as
\begin{align}
    E_{n_{+},\ell}&=\hbar\sqrt{\omega_{c}^{2}+4m_{G}\Omega^{2}\frac{(m_{G}-m^{*})}{m^{*2}}+4\omega_{c}\Omega\frac{(m_{G}-m^{*})}{m^{*}}}\left(n_{+}+\frac{1}{2}\right)\nonumber\\ &-\hbar\left[\frac{\sqrt{\omega_{c}^{2}+4m_{G}\Omega^{2}\frac{(m_{G}-m^{*})}{m^{*2}}+4\omega_{c}\Omega\frac{(m_{G}-m^{*})}{m^{*}}}}{2}-\frac{\omega_{c}}{2}-\frac{m_{G}\Omega}{m^{*}}\right]\ell,\label{nmasimenos}
\end{align}
where $ 2n+|l| = n_{+} + n_{-}$ and $\ell = n_{+} - n_{-}$, with $ n_{\pm} = 0, 1, 2, 3, \ldots $(see appendix \ref{A}).

We can rewrite the above expression as follows:
\begin{eqnarray}
    E_{n,\ell}=\hbar\omega(n_{+}+\frac{1}{2})-\hbar\varpi\ell, \label{eq:4.8}
\end{eqnarray}
where 
\begin{align*}
&\omega\equiv\sqrt{\omega_{c}^{2}+4m_{G}\Omega^{2}\frac{(m_{G}-m^{*})}{m^{*2}}+4\omega_{c}\Omega\frac{(m_{G}-m^{*})}{m^{*}}},\notag\\
&\varpi\equiv\frac{\sqrt{\omega_{c}^{2}+4m_{G}\Omega^{2}\frac{(m_{G}-m^{*})}{m^{*2}}+4\omega_{c}\Omega\frac{(m_{G}-m^{*})}{m^{*}}}}{2}-\frac{\omega_{c}}{2}-\frac{m_{G}\Omega}{m^{*}}.    
\end{align*}

To ensure that the energy eigenvalues acquire positive values, it is necessary that $\varpi \ell<0$, which means that we must have either $\varpi<0$ with $\ell>0$ or $\varpi>0$ with $\ell<0$. The first option occurs for $\Omega>0$, while the second holds for $\Omega<0$, therefore
$\varpi=\pm|\varpi|$ and $\ell=\mp|\ell|$, so that Eq. (\ref{eq:4.8}) can be rewritten as
\begin{eqnarray}
    E_{n,\ell}=\hbar\omega(n_{+}+\frac{1}{2})+\hbar|\varpi||\ell|.
\end{eqnarray}
Substituting $|\varpi|=\Gamma$ and $|\ell|=l$, we obtain
\begin{eqnarray}
    E_{n_+,l}=\hbar\omega(n_{+}+\frac{1}{2})+\hbar\Gamma l.\label{eq:4.10}
\end{eqnarray}

Setting $\Omega \equiv 0$, it yields $omega\equiv\omega_c$ and $\Gamma\equiv0$. This way, we have
\begin{eqnarray}
 E_{n_{+}} = \hbar \omega_{c} \left( n_{+} + \frac{1}{2} \right) (\text{Landau levels} ).
\end{eqnarray}
Taking $ m_{G} \equiv m^{*}$ \cite{BRANDAO201555,Brandao2017} in Eq. (\ref{nmasimenos}), it yields $\omega\equiv\omega_c$ and $\Gamma\equiv\Omega$, so we arrive at
\begin{eqnarray}
    E_{n_+,l}=\hbar\omega_c(n_{+}+\frac{1}{2})+\hbar\Omega l, \label{eq:4.10}
\end{eqnarray}
which is equivalent to
\begin{eqnarray}
  E_{n \ell} = \hbar \omega_{c} \left[ \left( n + \frac{|\ell| + \ell}{2} \right) + \frac{1}{2} \right] + \hbar \Omega \ell. \label{old}
\end{eqnarray}
If we take $\omega_c\equiv0$ in Eq. (\ref{nmasimenos}), then we recover the energy spectrum obtained in Ref.  \cite{WANG2024107117}, that is, 
\begin{eqnarray}
  E_{n_{+}n_{-}}&=&\hbar\Omega\left[\sqrt{m_{G}\frac{\left(m_{G}-m^{*}\right)}{m^{*2}}}\right]\left(n_{+}+n_{-}+1\right)+\hbar\Omega(n_{+}-n_{-})\nonumber\\ &=&2\hbar\Omega\left[\sqrt{m_{G}\frac{\left(m_{G}-m^{*}\right)}{m^{*2}}}\right]\left(n+\frac{|l|}{2}+\frac{1}{2}\right)+\hbar\Omega l.
\end{eqnarray}
This last result shows that a rotating system without an applied magnetic field can yield Landau-like levels in this case, in contrast to the one where $m_G\equiv m^*$, for which \cite{10.1119/1.19503}
\begin{eqnarray}
    E_l(\lambda)=\frac{\hbar^{2}\lambda^{2}}{2m^{*}}+\hbar\Omega l,\label{omegazero}
\end{eqnarray}
where $\lambda$ is a continuous variable if the sample is infinite.

For $\omega_{c}\gg\Omega$ in Eq. (\ref{nmasimenos}), we have
\begin{eqnarray}
  E_{n_{+}n_{-}}&=&\hbar\left[\omega_{c}+\frac{m_{G}\Omega}{m^{*}}\right]\left(n_{+}+\frac{1}{2}\right)+\hbar\left[-\frac{m_{G}\Omega}{m^{*}}\right]\left(n_{-}+\frac{1}{2}\right)\nonumber\\ &=&\hbar\omega_{c}(n_{+}+\frac{1}{2})+\hbar\frac{m_{G}\Omega}{m^{*}}(n_{+}-n_{-})\nonumber\\ &=&\hbar\omega_{c}\left(n+\frac{|l|+l}{2}+\frac{1}{2}\right)+\hbar\frac{m_{G}}{m^{*}}\Omega l,  
\end{eqnarray}
which is the same as Eq. (\ref{old}), apart from the term $m_G/m^*$.

The energy structure for the 2DEG obtained here directly impacts the thermodynamic properties, as we will see in the following section.

\FloatBarrier
\section{Partition function and Thermodynamic Properties}\label{sec3}

\subsection{Partition function}

Although the system is a two-dimensional electron gas (2DEG), the use of the Boltzmann partition function is justified due to the statistical distribution of energy states, which are modified by rotation and the magnetic field. However, it can still be treated within the canonical formulation of statistical mechanics. Furthermore, the approach assumes that electron interactions are negligible, making applying the Boltzmann distribution valid. It is also worth noting that, in this study, we consider moderate magnetic fields and rotations, avoiding extreme regimes that could require a Fermi-Dirac statistical formulation. 

The partition function will be given by
\begin{equation}
	Z=\sum_{n_{+},l}e^{-\frac{E_{n_{+},l}}{k_BT}}=\sum_{n_{+}=0}^{\infty}e^{-\frac{E_{n_{+}}}{k_BT}}\sum_{l=0}^{\infty}e^{-\frac{E_{l}}{k_BT}}.
\end{equation}
Considering the energy spectrum (\ref{nmasimenos}), it yields
\begin{eqnarray}
	Z=\frac{e^{\frac{\hbar\varGamma}{2k_BT}}}{4\sinh(\frac{\hbar\omega}{2k_BT})\sinh(\frac{\hbar\Gamma}{2k_BT})},
\end{eqnarray}
where $\omega$ and $\Gamma$ are given as shown above.

In what follows, when the non-rotating case is mentioned, we use the partition function for the degenerate Landau levels.
\begin{equation}
Z_L = \frac{m^* \omega_c A}{4 \pi \hbar} \, \csch\left( \frac{\hbar \omega_c}{2k_BT} \right),
\end{equation}
where $ m^* $ is the effective mass of the electron, $ \omega_c $ is the frequency associated with the magnetic field and $ A =1.0$ mm$^2$ is the area of the system. We consider $m^*\approx0.067\,m_e$ (GaAs), with $m_e\approx9.109 \times 10^{-31}~$Kg. $ T $ is the absolute temperature.

\subsection{Thermodynamic Properties}

Using the partition function, various thermodynamic quantities can be determined through the following relations:

\subsubsection{Mean Energy}
The system's mean (or internal) energy $U$ is derived from the partition function $Z$ logarithm. It represents the average energy of the system's microstates weighted by their Boltzmann probability. It is given by
\begin{equation}
U = -\left(\frac{\partial \ln Z}{\partial \beta}\right)_{B,\Omega}\;,
\end{equation}
with $\beta = 1/k_B T$, where $k_B$ is the Boltzmann constant. 

\subsubsection{Heat Capacity}
The heat capacity at constant volume, $C_V$, measures how much energy is required to raise the system's temperature. It is the derivative of the mean energy with respect to temperature, showing how the internal energy changes as the system's temperature changes. Here, it is equivalently calculated while keeping both $B$ and $\Omega$ constant, that is,  
\begin{equation}
C_{B,\Omega} = \left(\frac{\partial U}{\partial T}\right)_{B,\Omega} = k_B \beta^2 \left(\frac{\partial^2 \ln Z}{\partial \beta^2 }\right)_{B,\Omega}.
\end{equation}

\subsubsection{Free Energy}
The free energy $F$ represents the work that can be extracted from the system at a given temperature. It combines internal energy and entropy, providing insight into the system's stability and equilibrium properties. It is derived directly from the partition function:
\begin{equation}
F = -k_B T \ln Z.
\end{equation}

\subsubsection{Entropy}
The entropy $ S $ measures the system's disorder or randomness. It is derived from the partition function and related to the number of accessible microstates. Entropy is crucial for understanding the flow of energy and the distribution of particles within the system. It is given by
\begin{equation}
    S(T,B) = k_B T \left(\frac{\partial \ln Z}{\partial T}\right)_{B,\Omega}.
\end{equation}

\subsubsection{Magnetization}

The magnetization of a 2DEG takes into account the quantization of energy levels in a magnetic field, and the influence of thermal effects, and it is defined as
\begin{equation} 
M(T,B) = T\, \left(\frac{\partial \ln Z}{\partial B}\right)_{T,\Omega},
\end{equation}
where $ Z $ is the partition function. 

\subsubsection{Magnetocaloric Effect}

The magnetocaloric effect in a 2DEG is obtained from the entropy $S(T, B, \Omega)$, where $ T $ is the temperature, $ B $ is the magnetic field, and $ \Omega $ is the system's rotation. To analyze the effect at {\it constant} rotation ($d\Omega\equiv 0$), consider the isoentropic condition, $ dS = 0 $, which implies
\begin{equation}
    dS = \left( \frac{\partial S}{\partial T} \right)_{B} dT + \left( \frac{\partial S}{\partial B} \right)_{T} dB = 0.
\end{equation}

Isolating $dT/dB$, we obtain
\begin{equation}
    \frac{dT}{dB} = -\frac{\left( \frac{\partial S}{\partial B} \right)_{T}}{\left( \frac{\partial S}{\partial T} \right)_{B}},
\end{equation}
which describes the temperature variation with the magnetic field under isoentropic conditions. $B_i$ and $B_f$ denote the initial and final values of the magnetic field, respectively, over which the adiabatic process and the MCE are evaluated. From it, we have
\begin{eqnarray}
    \Delta T_{B}=-\intop_{B_{i}}^{B_{f}}\left[\frac{\left(\frac{\partial S}{\partial B}\right)_{T}}{\left(\frac{\partial S}{\partial T}\right)_{B}}\right]dB,
\end{eqnarray}
which is the adiabatic change in temperature of the magnetic system around temperature a $T$. Considering 
\begin{equation}
    \left( \frac{\partial M}{\partial T} \right)_B = - \left( \frac{\partial S}{\partial B} \right)_T\;
\end{equation}
and
\begin{equation}
    C_B = T\left( \frac{\partial S}{\partial T} \right)_B,
\end{equation}
the magnetocaloric effect can be rewritten as 
\begin{equation}
    \Delta T_{B} = - \int_{B_i}^{B_f} \frac{T}{ C_B}\left( \frac{\partial M}{\partial T} \right)_B dB.
\end{equation}

\section{Results and discussions}\label{sec4}

In all subsequent analyses, we consider the thermodynamic quantities mentioned above as functions of temperature ($K$) for magnetic fields $B=0$ T, $B=0.01$ T, $B=0.1$ T, and $B=0.5$ T. For each of these cases, curves corresponding to different values of the rotation frequency $\omega$ ($\pm 50\,\mathrm{GHz}$ and $\pm 100\,\mathrm{GHz}$) are presented alongside the non-rotating case ($\omega=0$). Initially, we will analyze the scenario where $ m^* \neq m_{G} $, followed by the case where $m^* \equiv m_{G}$. Note that the case $B=0$ T is not displayed for the latter scenario since it does not exhibit a Landau-level structure, as demonstrated in Eq.~\eqref{omegazero}.

\subsection{Internal Energy}

\begin{figure}[hbt!]
{\includegraphics[width=0.47\linewidth]{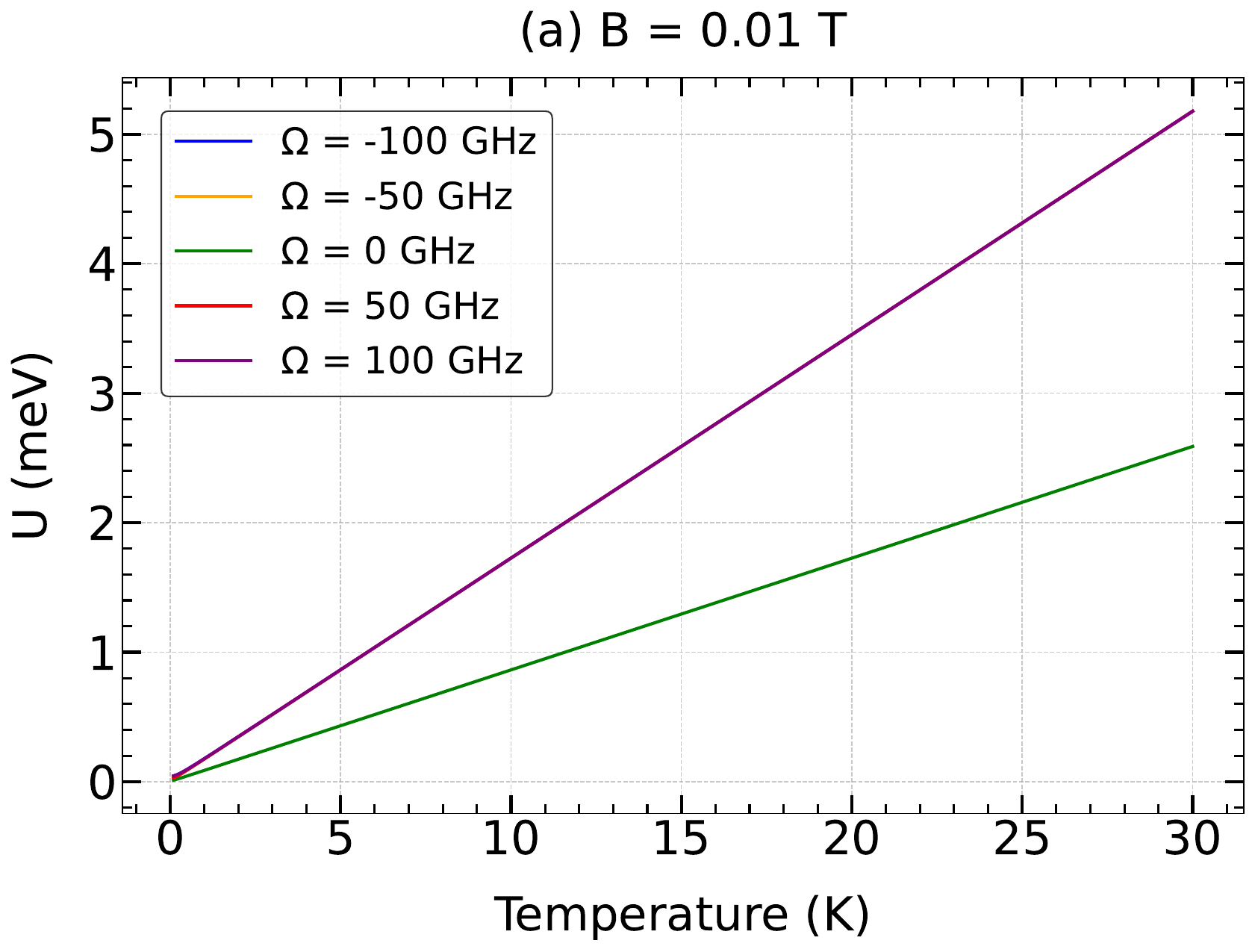}}\qquad \vspace{0.3cm}
{\includegraphics[width=0.47\linewidth]{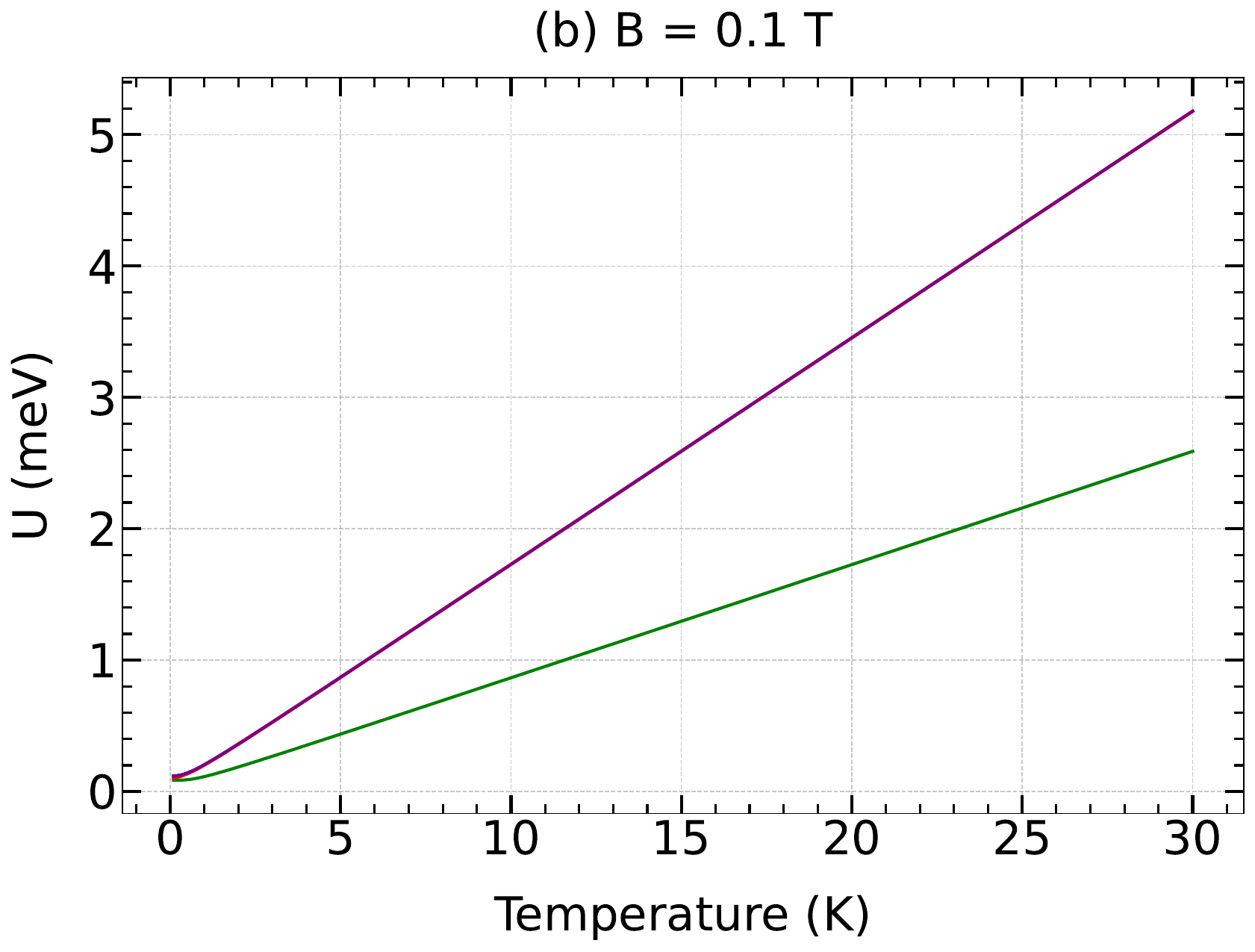}}
{\includegraphics[width=0.47\linewidth]{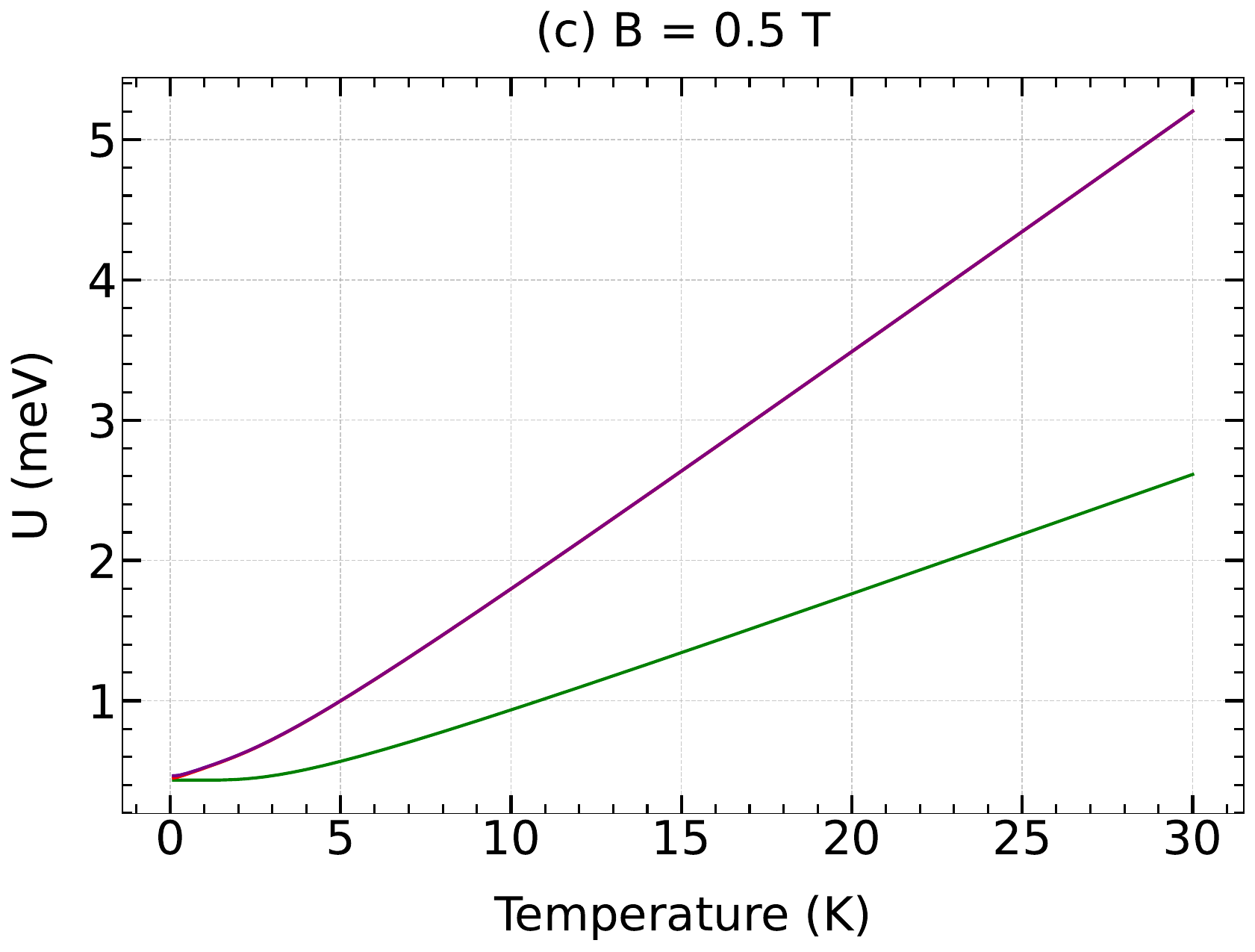}}
     \caption{Internal energy for a rotating 2DEG for some values of angular speed $\Omega$ as a function of temperature and for different values of an external magnetic field intensity (measured in Tesla). The plots where $\Omega = 0$ were computed for the problem of degenerate Landau levels. Here, $m^*\equiv m_G$.}
     \label{internalU}
\end{figure}

In the case of $B=0$ T (Fig. \ref{internalU}-(a)), there is no curve for the nonrotating case because, without a magnetic field, the system does not exhibit Landau levels for comparison. The curves show that the internal energy increases with temperature almost linearly, except at very low temperatures, with different slopes and absolute values. The curves corresponding to rotation frequencies of $\pm 50\,\mathrm{GHz}$ coincide with each other, as do the curves for $\pm 100\,\mathrm{GHz}$. The $\pm 100\,\mathrm{GHz}$ curves also show the most significant upward deviation.
\begin{figure}[t]
{\includegraphics[width=0.48\linewidth]{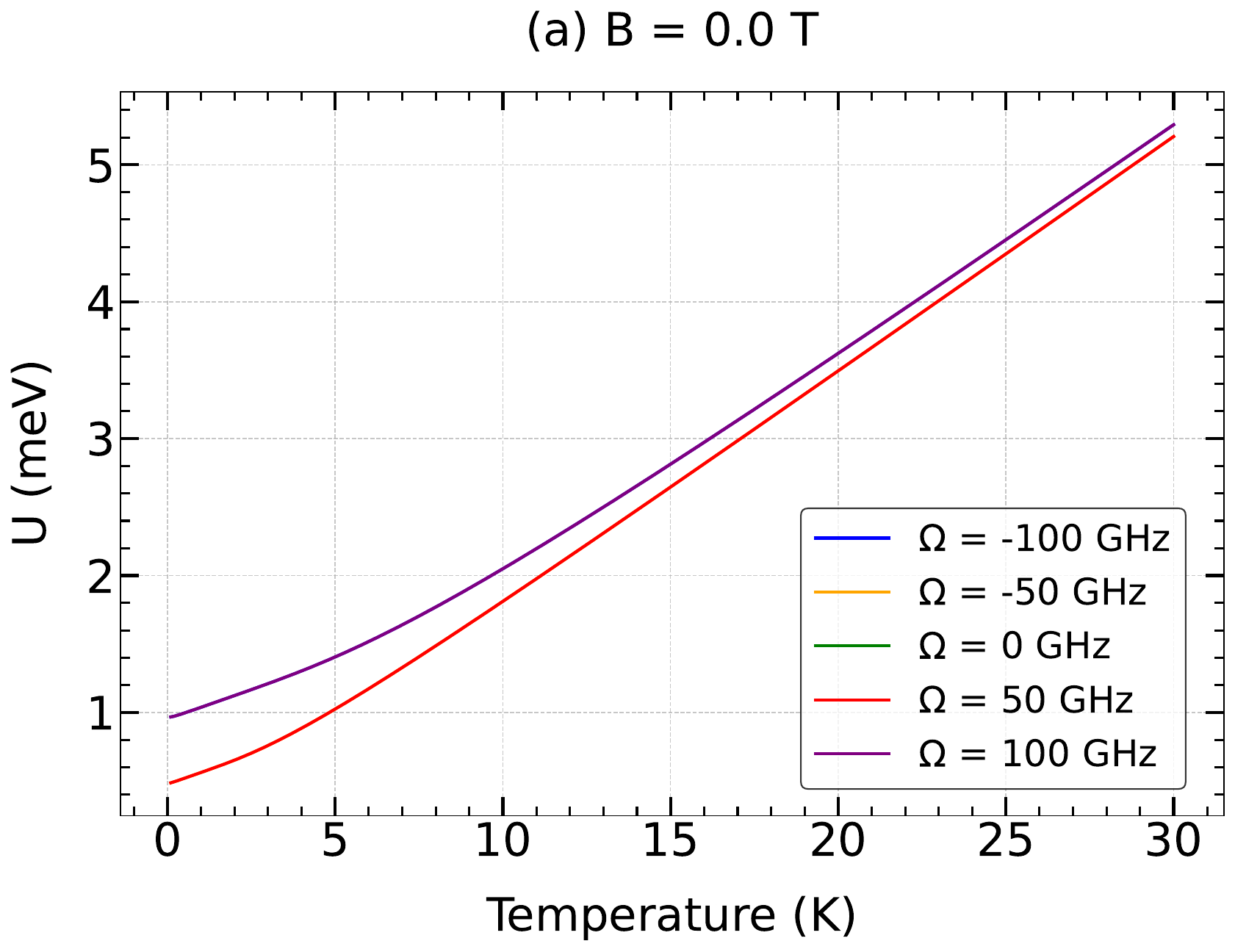}}\qquad \vspace{0.3cm}
{\includegraphics[width=0.48\linewidth]{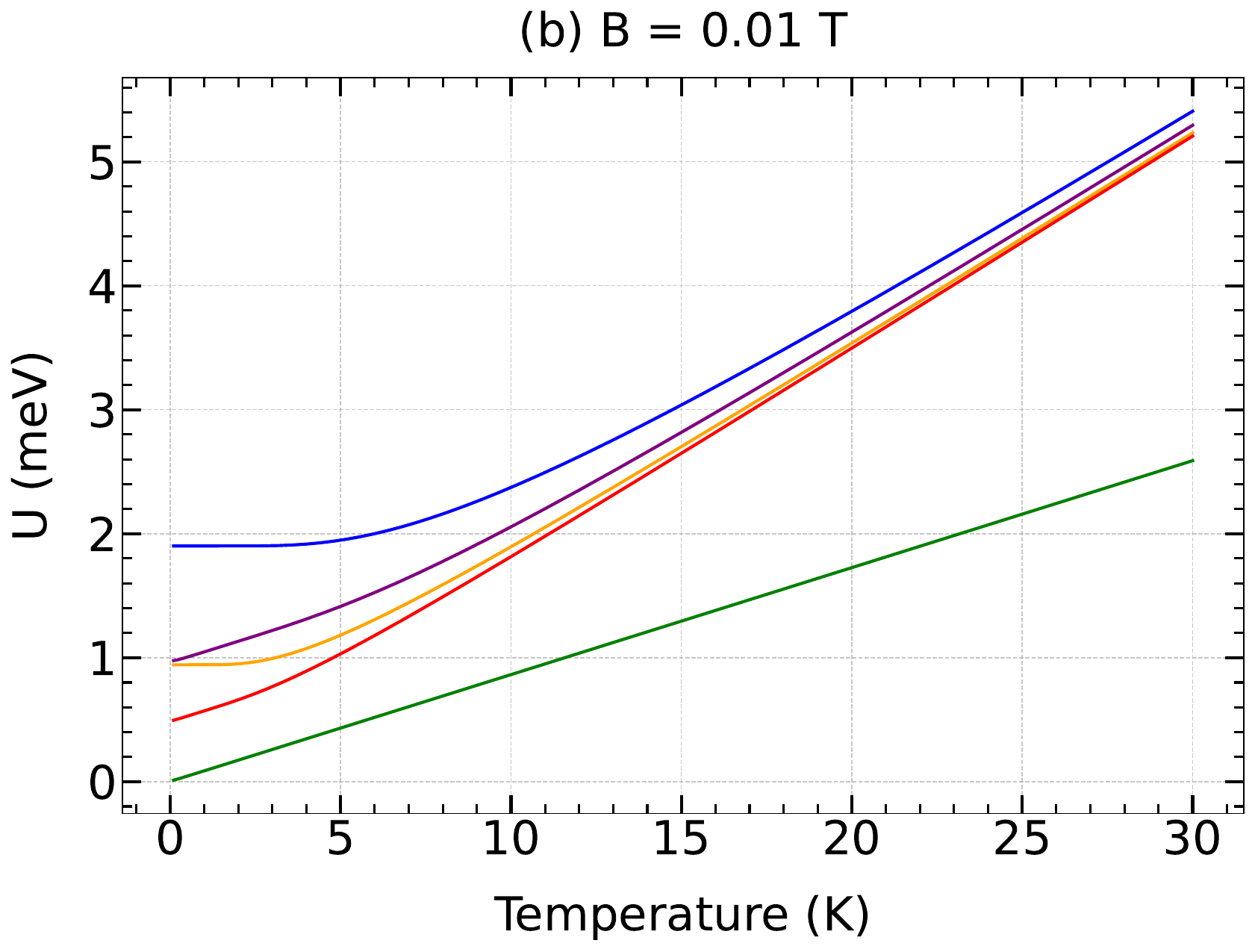}}
{\includegraphics[width=0.48\linewidth]{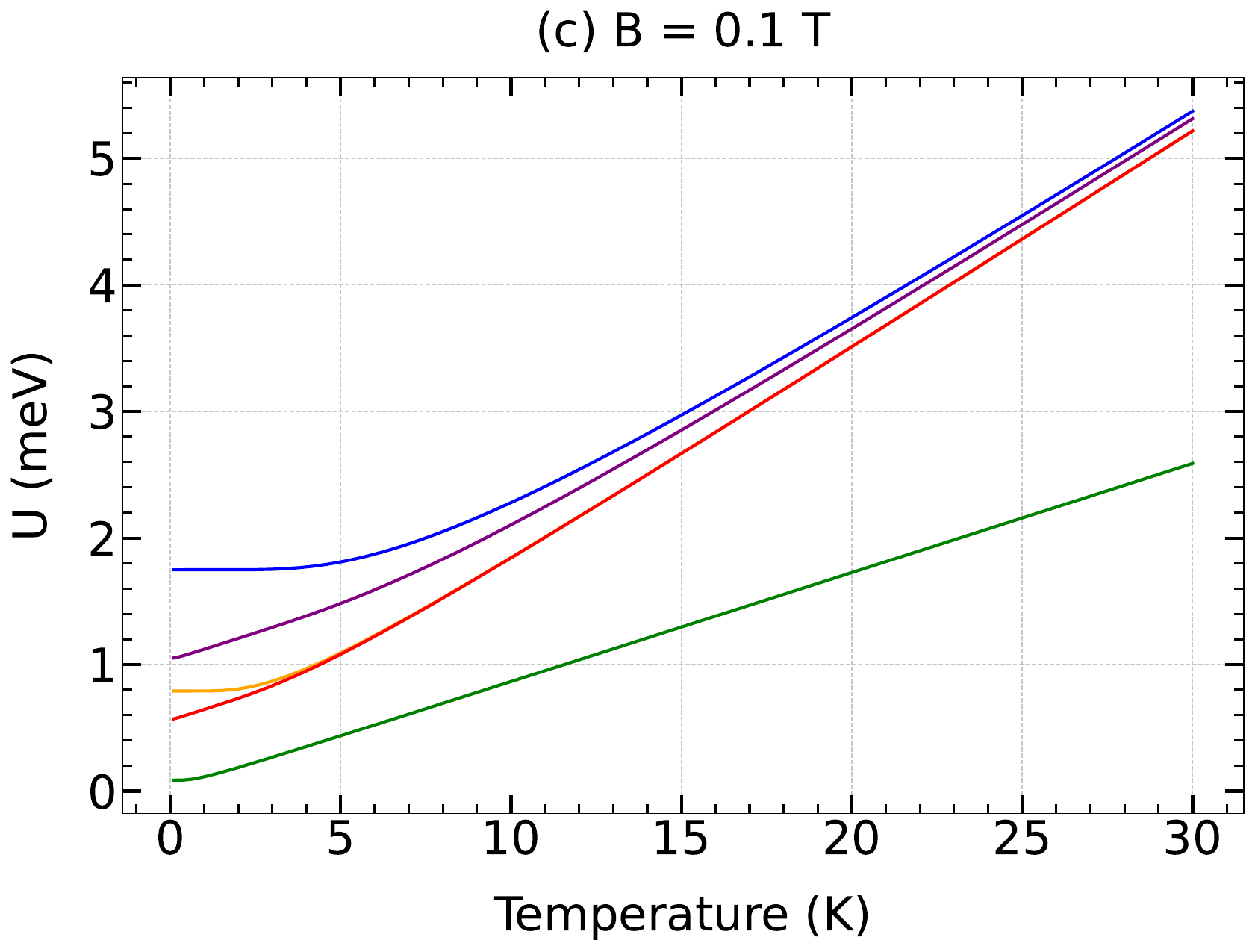}}\qquad
{\includegraphics[width=0.48\linewidth]{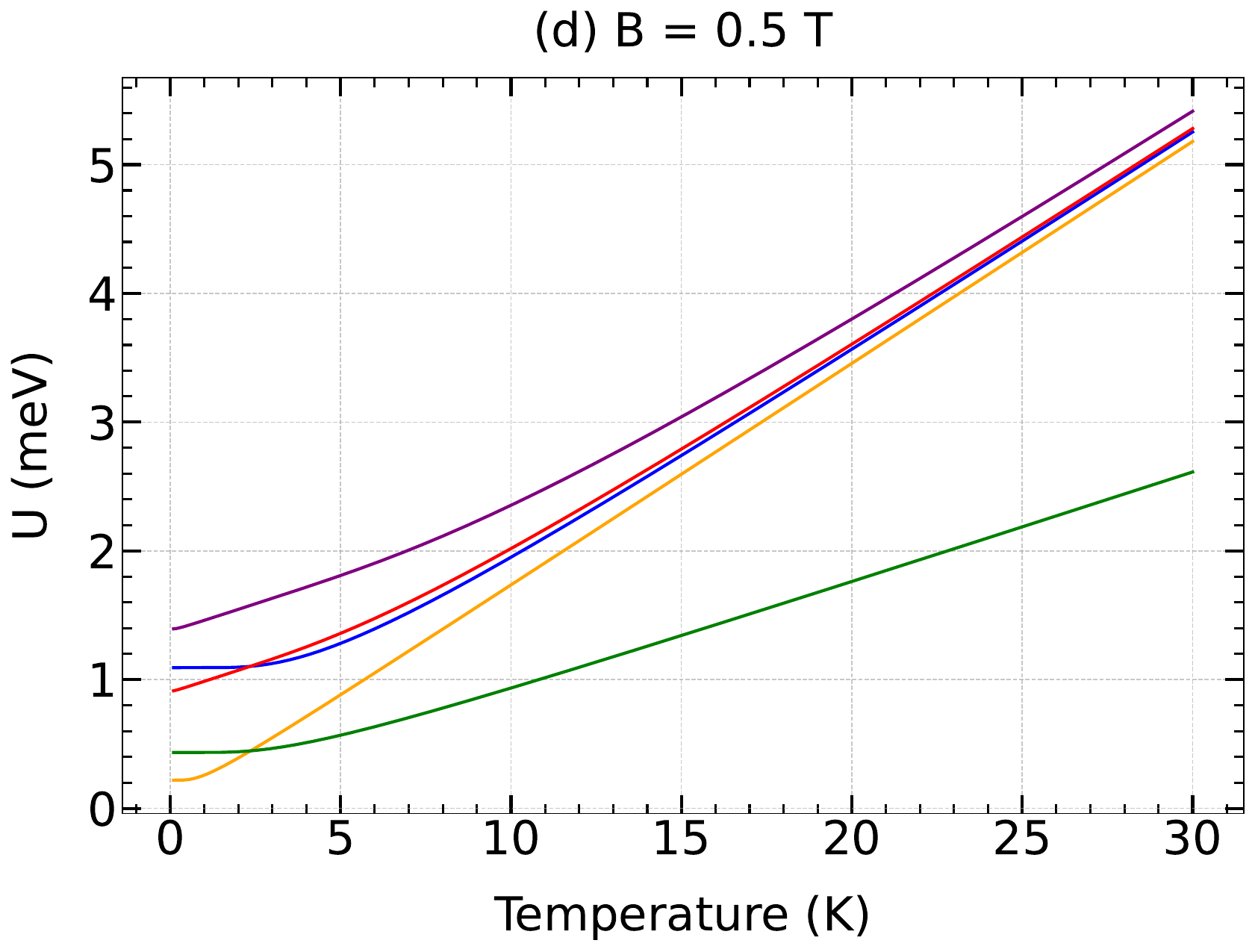}}
\caption{Internal energy for a rotating 2DEG for some values of angular speed $\Omega$ as a function of temperature and for different values of an external magnetic field intensity (measured in Tesla). The plots where $\Omega = 0$ were computed for the problem of degenerate Landau levels. Here, $m^*\neq m_G$.}
\label{internalUm}
\end{figure}
When introducing a nonzero magnetic field, the curves representing the non-rotating case appear. The Landau levels are highly degenerate, which is reflected in a distinct energy behavior. They generally remain at lower internal energy values than the rotating curves, especially as the temperature increases. For $B=0.01$ T (Fig.~\ref{internalUm}-(b))  $B=0.1$ T (Fig.~\ref{internalUm}-(c)), the rotation with $-100\,\mathrm{GHz}$ tends to lie above the others, followed by the rotation of the same magnitude (positive $+100\,\mathrm{GHz}$). Meanwhile, the $\pm 50\,\mathrm{GHz}$ rotations exhibit lower internal energy values. In fact, the behavior of the curves is similar to the $B=0\,\mathrm{T}$ case. For $B=0.5$ T (Fig. \ref{internalU}-(d)), the hierarchy among the curves is modified compared to the previous cases.

On the other hand, Fig.~\ref{internalU} considers the case where $ m^* \equiv m_{G} $, showing that the rotation $ \Omega $ has a significant impact on the internal energy; however, noticeable deviations between curves corresponding to different rotations considered are not observed. 

In general, it can be concluded that applying a magnetic field increases the system's internal energy and accentuates the differences between the rotating and non-rotating cases. The non-rotating curve, typical of highly degenerate Landau levels, usually exhibits $U$ values lower than those obtained when rotation is added to the system. The increase in temperature leads to a rise in $U$ in all scenarios.
\FloatBarrier
\subsection{Specific Heat}
\begin{figure}[t]
{\includegraphics[width=0.48\linewidth]{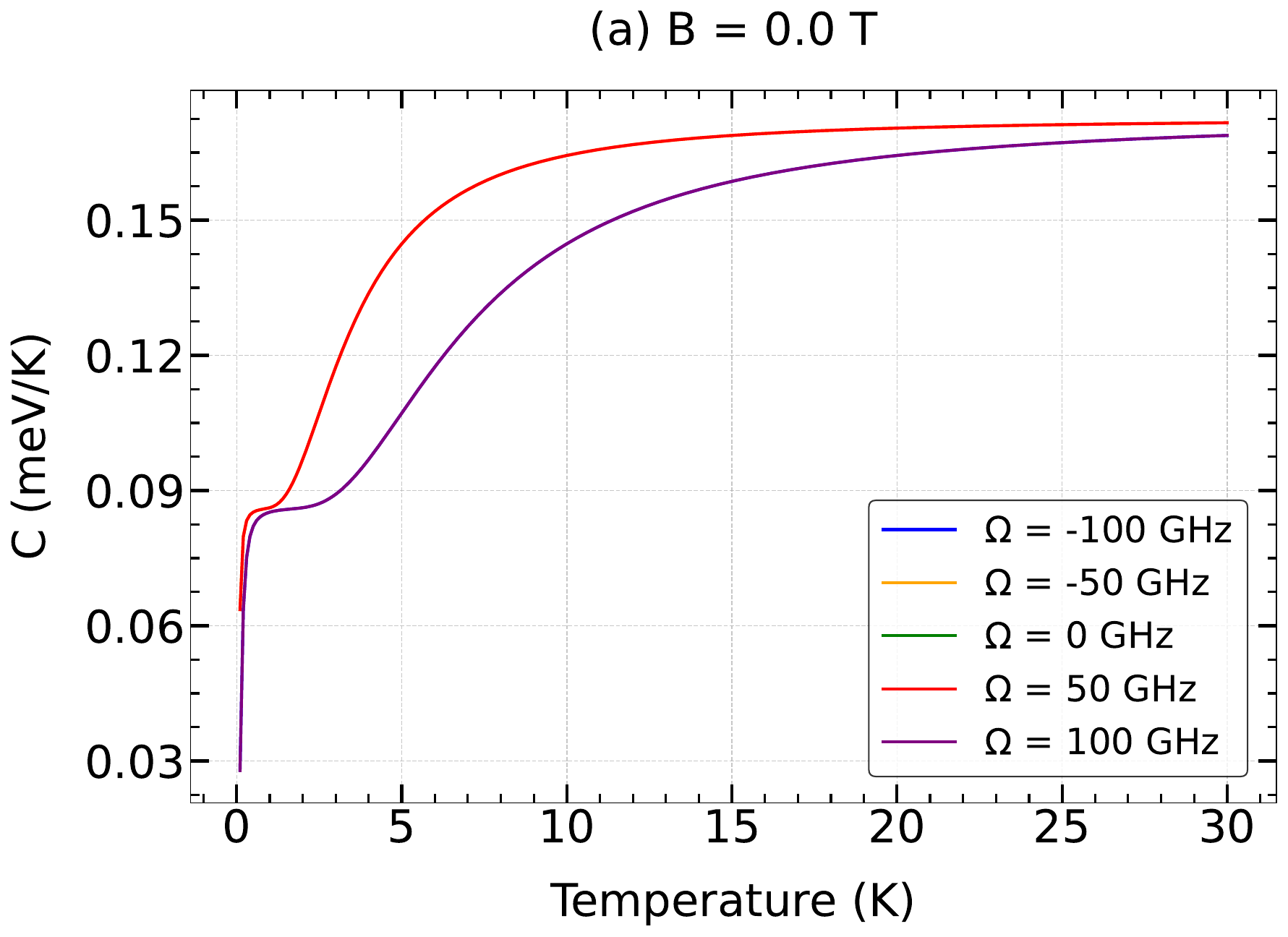}}\qquad \vspace{0.3cm}
{\includegraphics[width=0.48\linewidth]{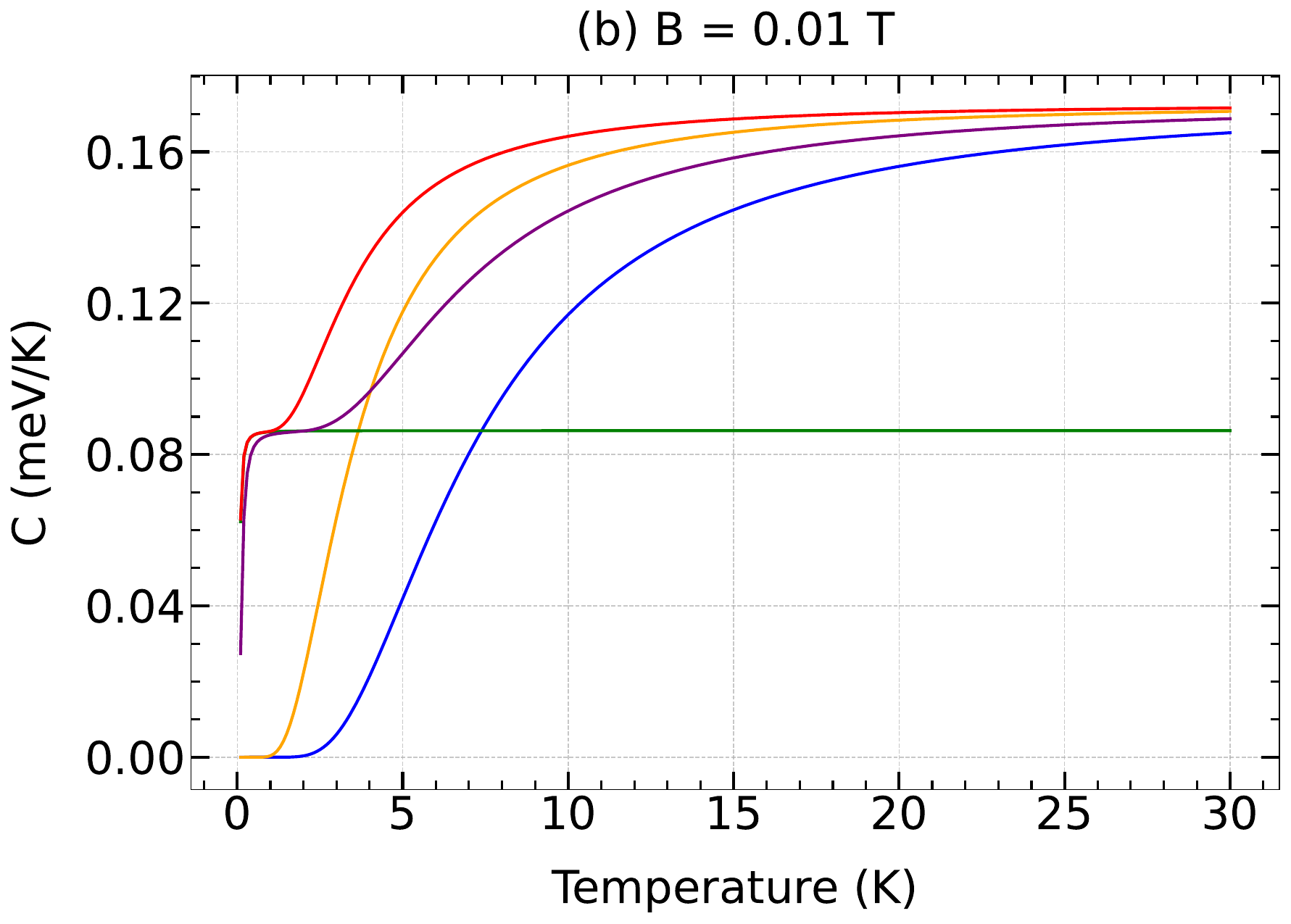}}
{\includegraphics[width=0.48\linewidth]{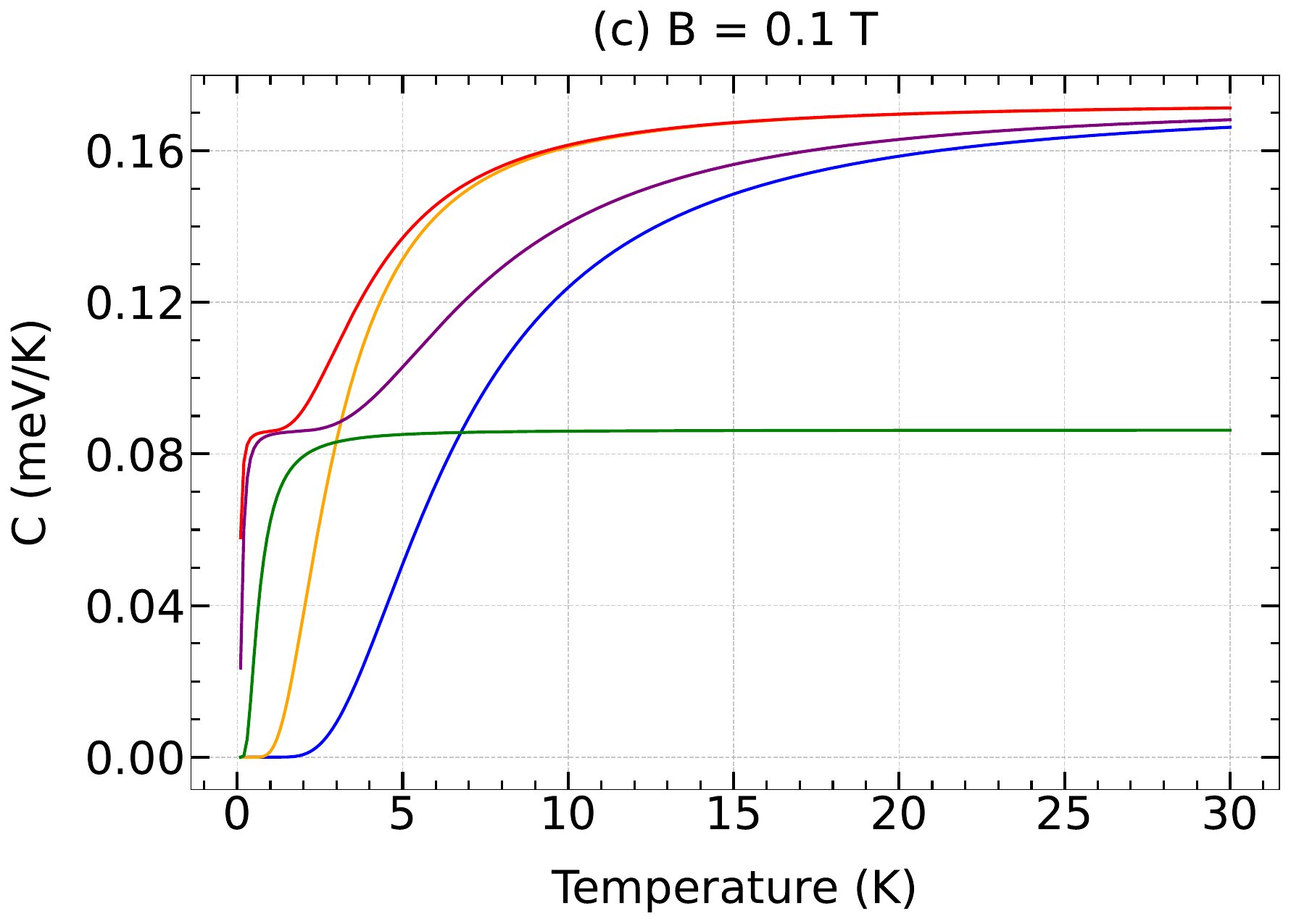}}\qquad
{\includegraphics[width=0.48\linewidth]{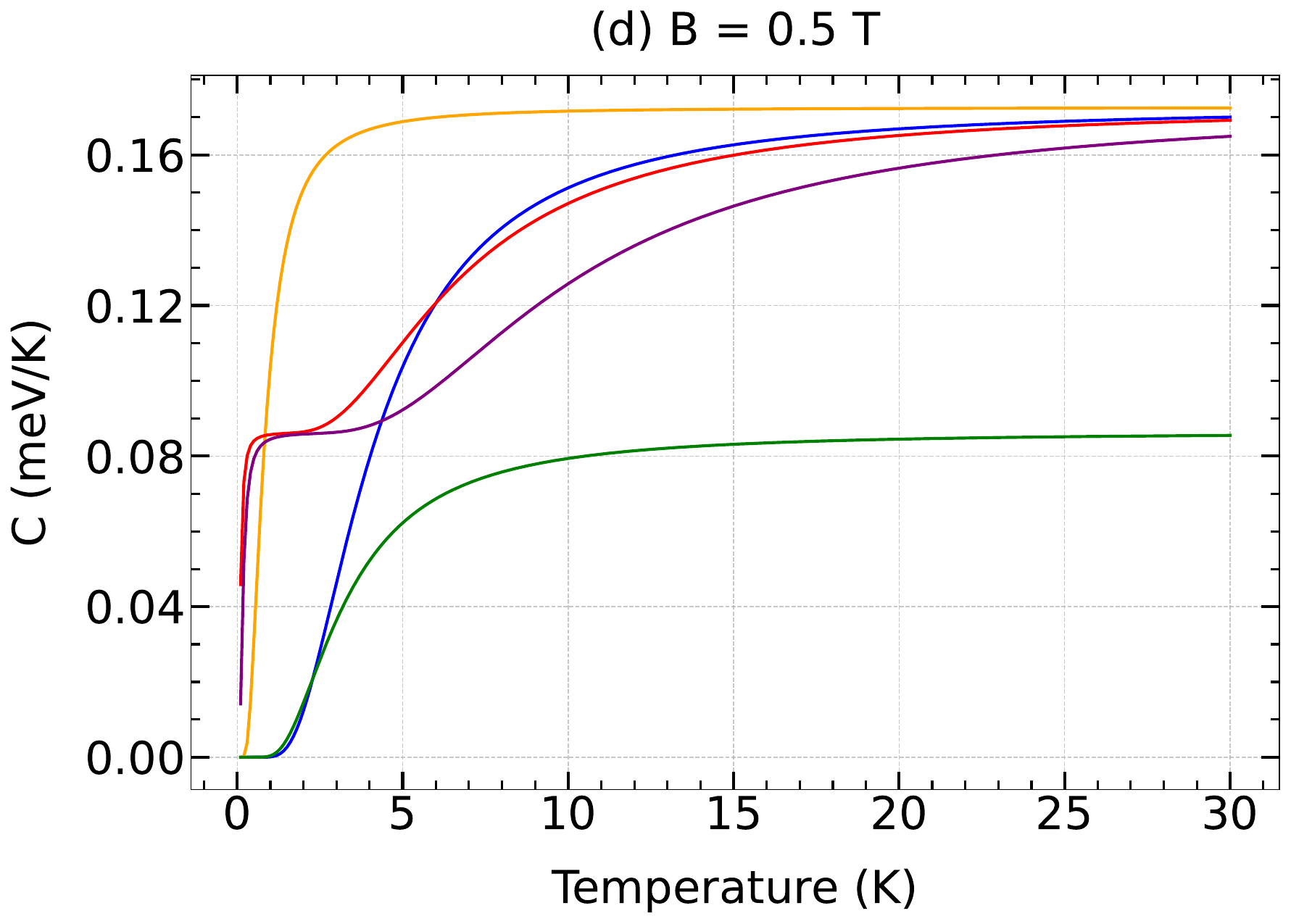}}
\caption{Specific heat for a rotating 2DEG for some values of angular speed $\Omega$ as a function of temperature and for different values of an external magnetic field intensity (measured in Tesla). The plots where $\Omega = 0$ were computed for the problem of degenerate Landau levels. Here, $m^*\neq m_G$.}
\label{specificheatm}
\end{figure}

From Fig.~\ref{specificheatm}-(a) ($B=0$ T), it can be observed that all curves start at $T=0$ K and increase rapidly as the temperature rises, approaching a plateau at temperatures above approximately $10$--$15\,\mathrm{K}$. The curves corresponding to rotation frequencies of $\pm 50\,\mathrm{GHz}$ coincide with each other, as do the curves for $\pm 100\,\mathrm{GHz}$. Lower rotation frequencies (in absolute value) result in higher $C$ values. 

In Figs.~\ref{specificheatm}-(b) and \ref{specificheatm}-(c), with the introduction of a magnetic field, the non-rotating curve appears, representing a system with Landau levels. It generally remains at lower $C$ values for intermediate and high temperatures compared to the rotating curves. The hierarchy between the $100\,\mathrm{GHz}$ and $-50\,\mathrm{GHz}$ curves compared to the $\Omega=0$ GHz one is observed at low temperatures. No caso $B=0.5$ T, Fig.~\ref{specificheatm}-(d), the effect of a stronger magnetic field becomes more evident, especially at low temperatures. There is a certain change in the hierarchy between all the curves. This phenomenon occurs because the magnetic field, together with the rotation, redistributes the energy levels and their densities of states.

\begin{figure}[t]
{\includegraphics[width=0.48\linewidth]{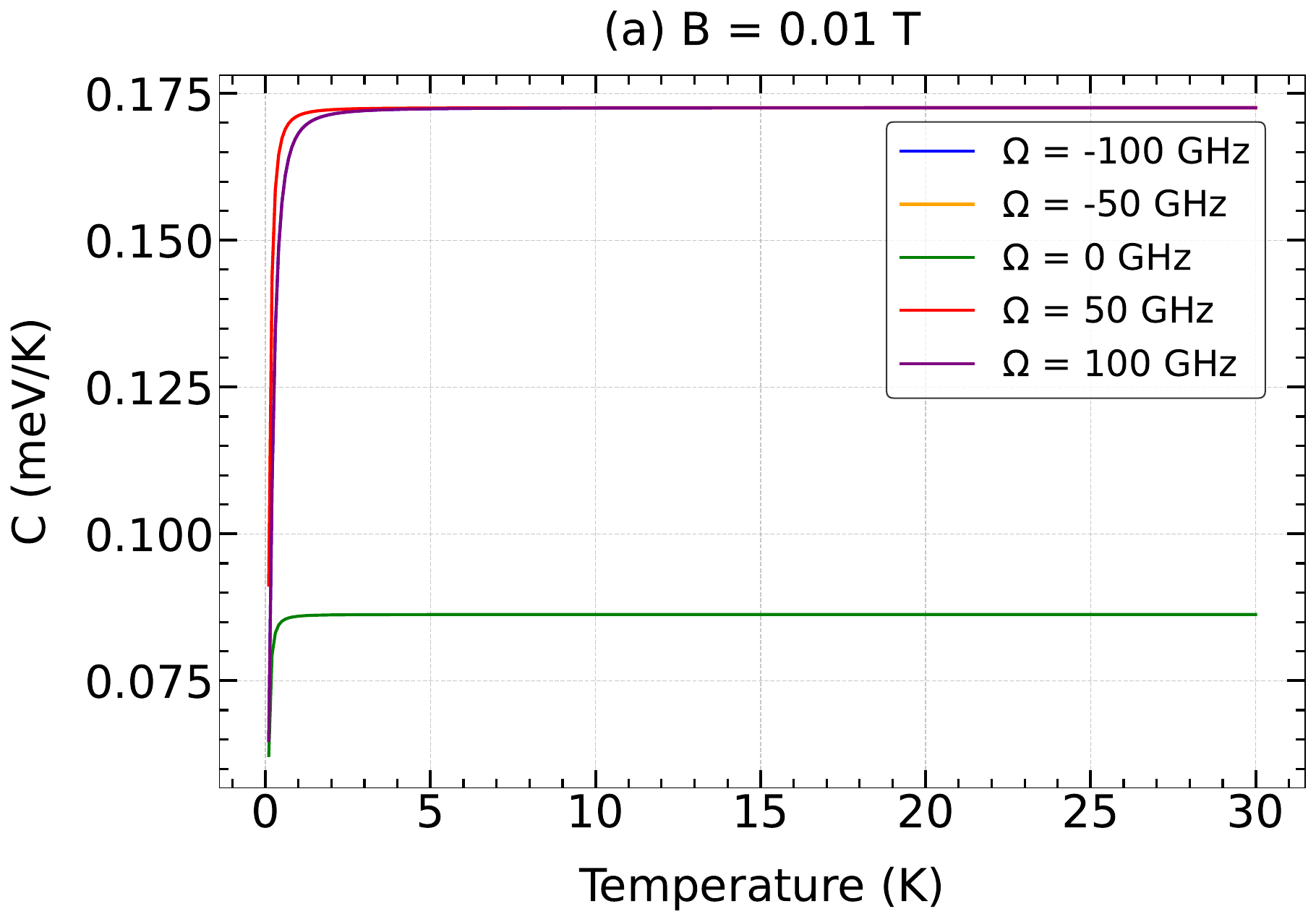}}\qquad \vspace{0.3cm}
{\includegraphics[width=0.48\linewidth]{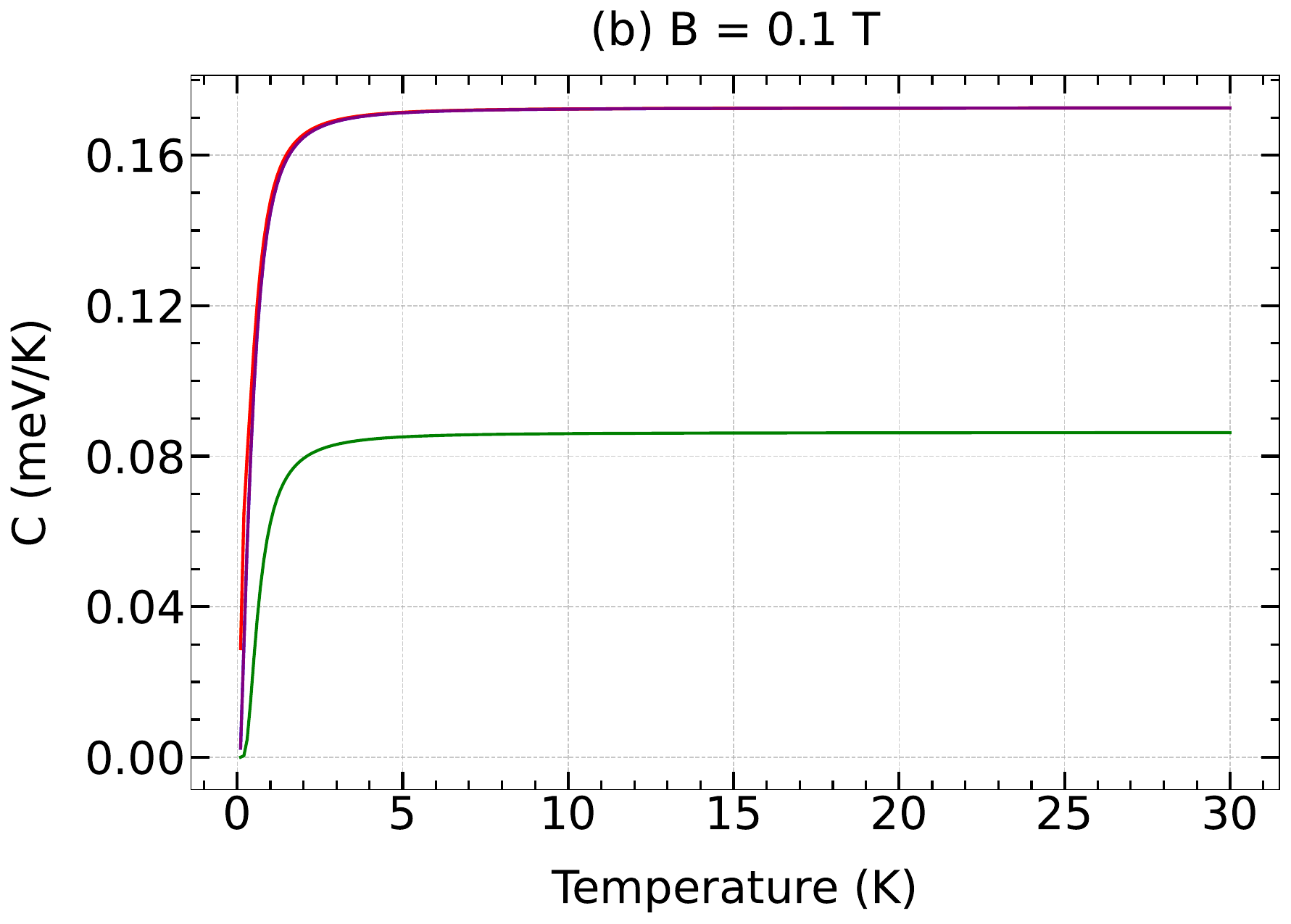}}
{\includegraphics[width=0.48\linewidth]{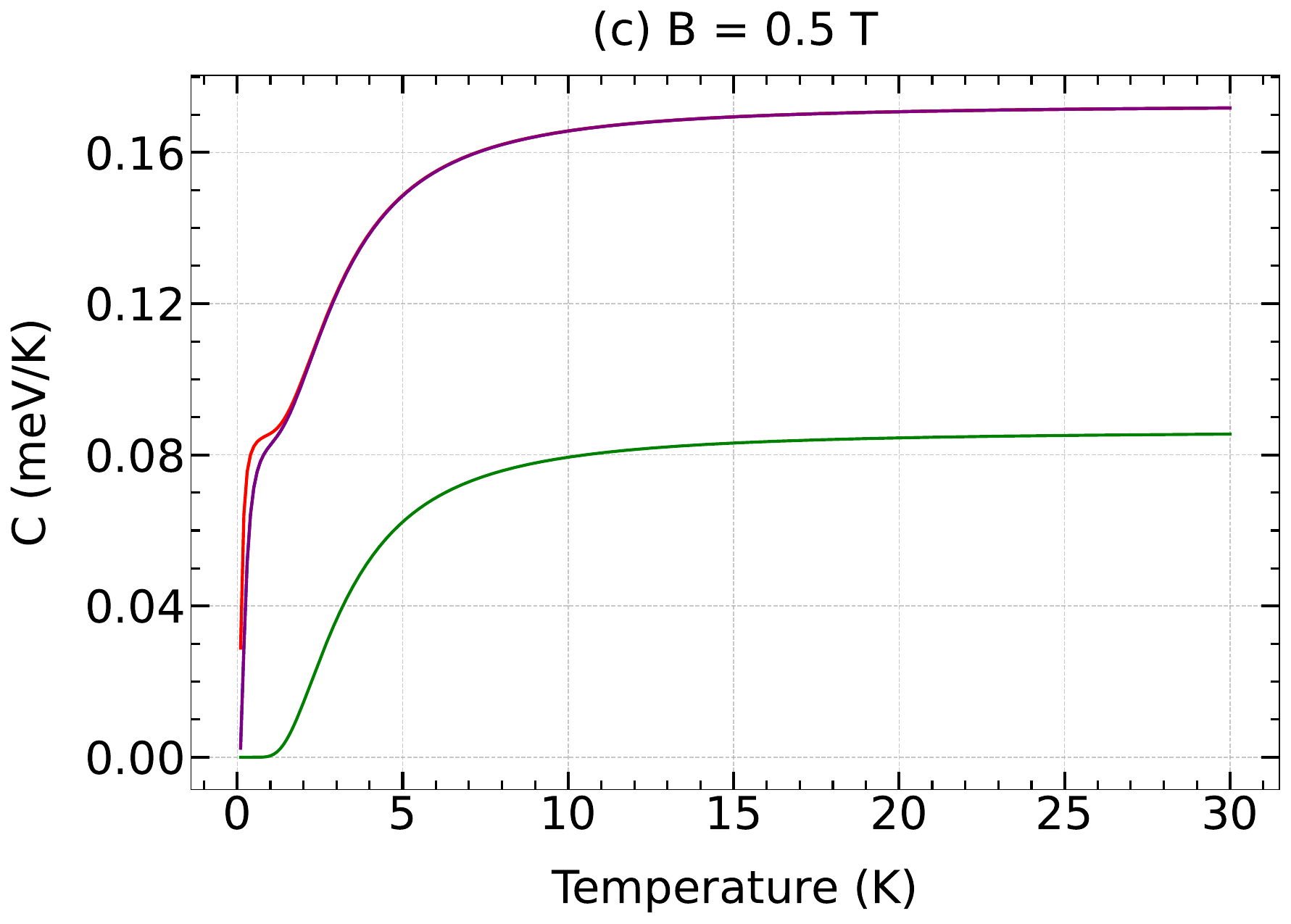}}
\caption{Specific heat for a rotating 2DEG for some values of angular speed $\Omega$ as a function of temperature and for different values of an external magnetic field intensity (measured in Tesla). The plots where $\Omega = 0$ Hz were computed for the problem of degenerate Landau levels. Here, $m^*\equiv m_G$.}
\label{specificheat}
\end{figure}

On the other hand, Fig.~\ref{specificheat} considers the case where $ m^* \equiv m_{G} $, showing that the rotation $\Omega$ has a significant impact on the specific heat as in the case for the internal energy; again, deviations between curves corresponding to different rotations considered are not significant.

In general, the specific heat tends to zero as $T \to 0$, in accordance with the Third Law of Thermodynamics. At intermediate and high temperatures, the value of $C$ reflects the number of accessible excited states, which is influenced by both the magnetic field and rotation. While the magnetic field alone produces highly degenerate Landau levels, rotation breaks or modifies this degeneracy, resulting in a greater density of possible transitions and, consequently, a higher specific heat. The rotation sign also affects the spectrum's shape in the presence of specific values of $B\neq 0$, potentially changing the hierarchy among the curves in different temperature ranges.

\FloatBarrier
\subsection{Free Energy}

\begin{figure}[t!]
{\includegraphics[width=0.48\linewidth]{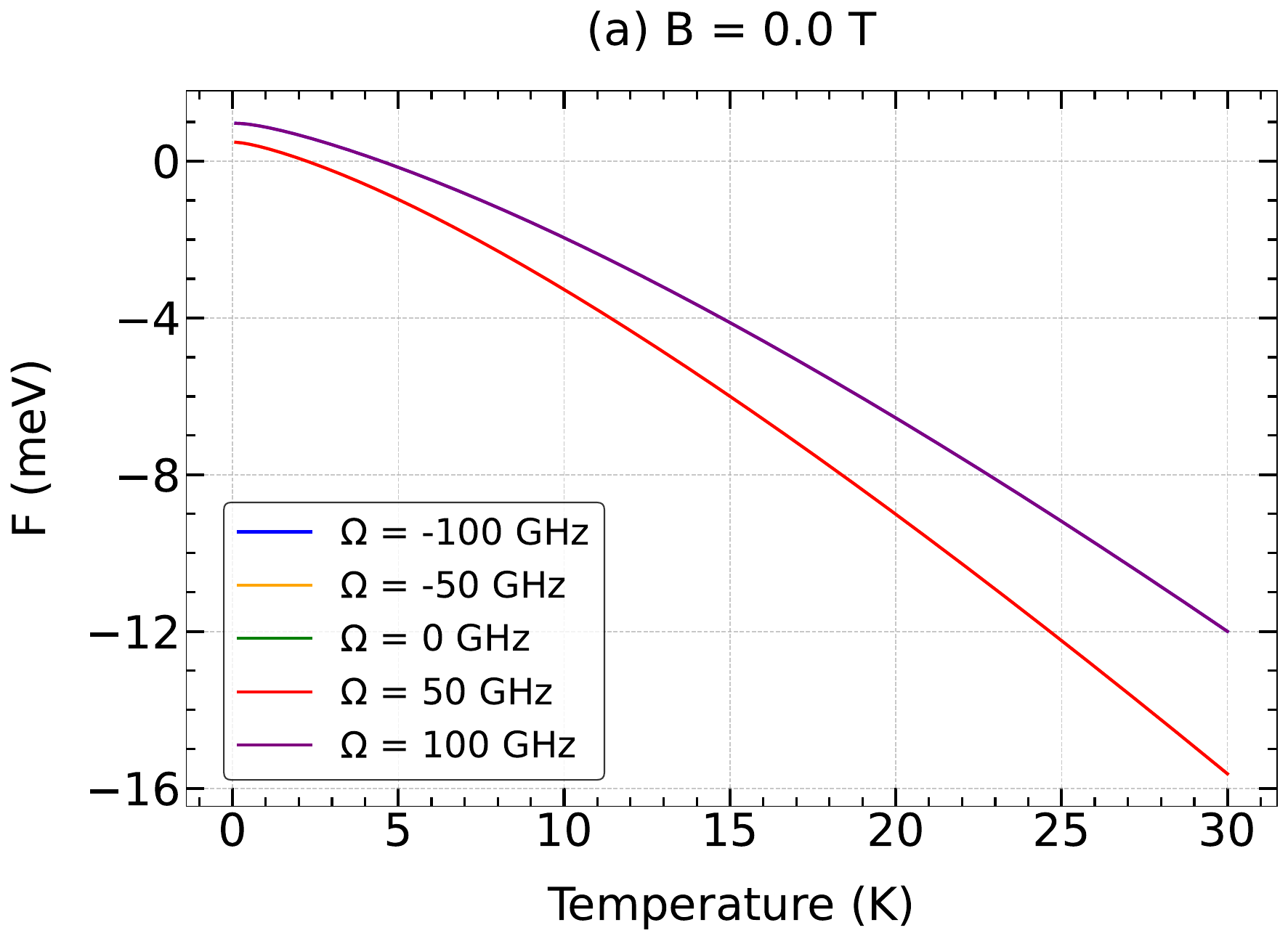}}\qquad \vspace{0.3cm}
{\includegraphics[width=0.48\linewidth]{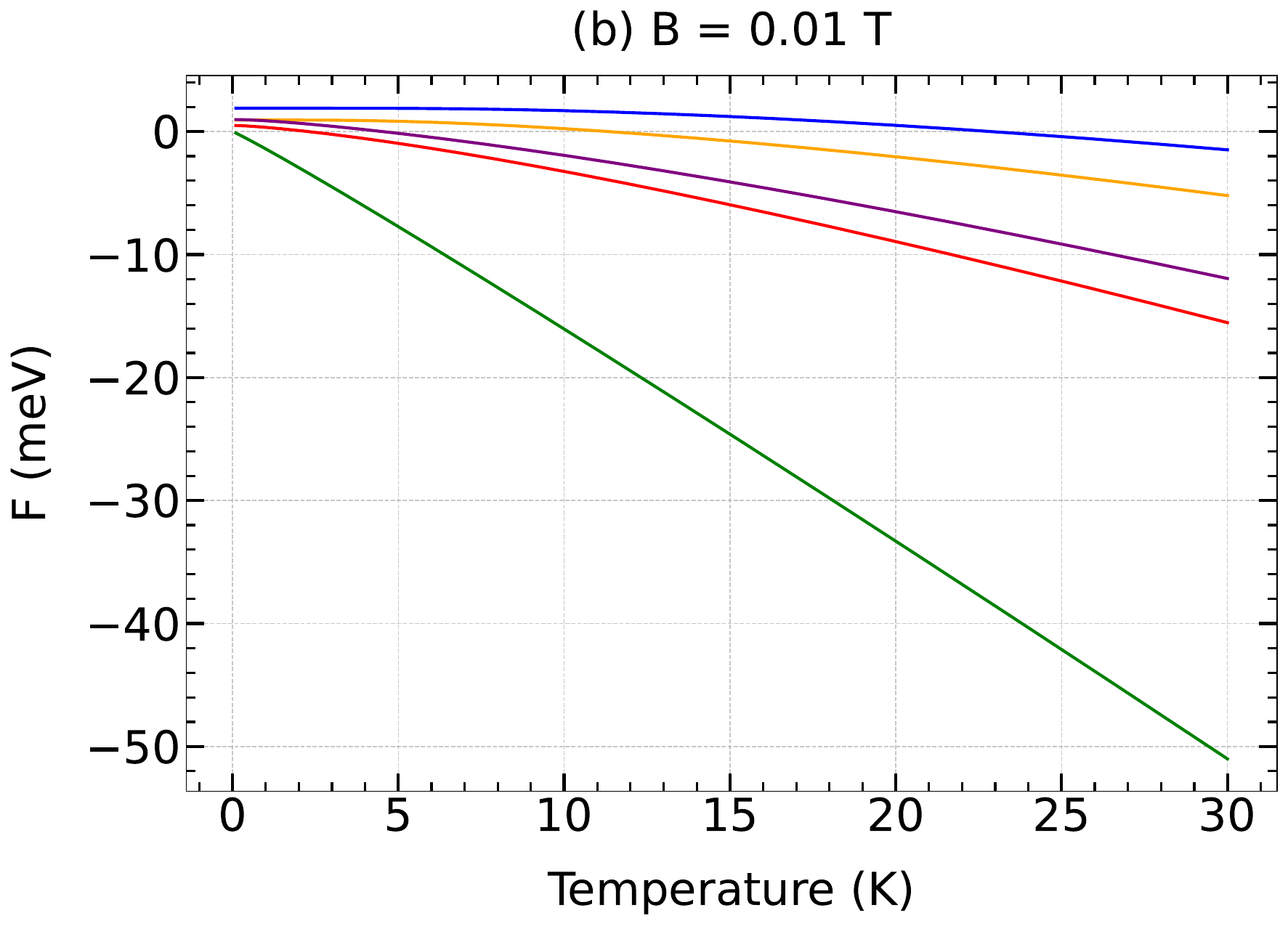}}
{\includegraphics[width=0.48\linewidth]{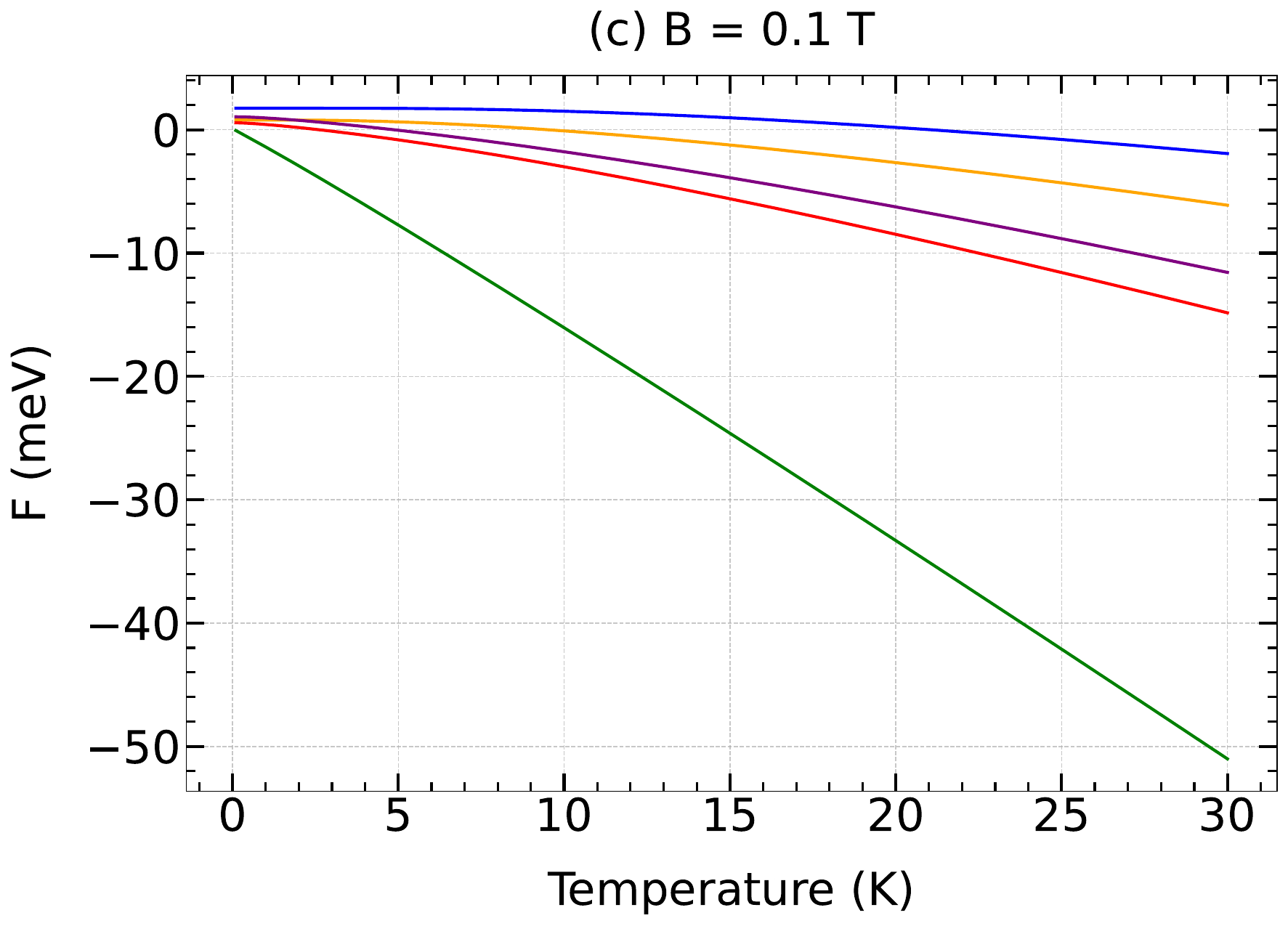}}\qquad
{\includegraphics[width=0.48\linewidth]{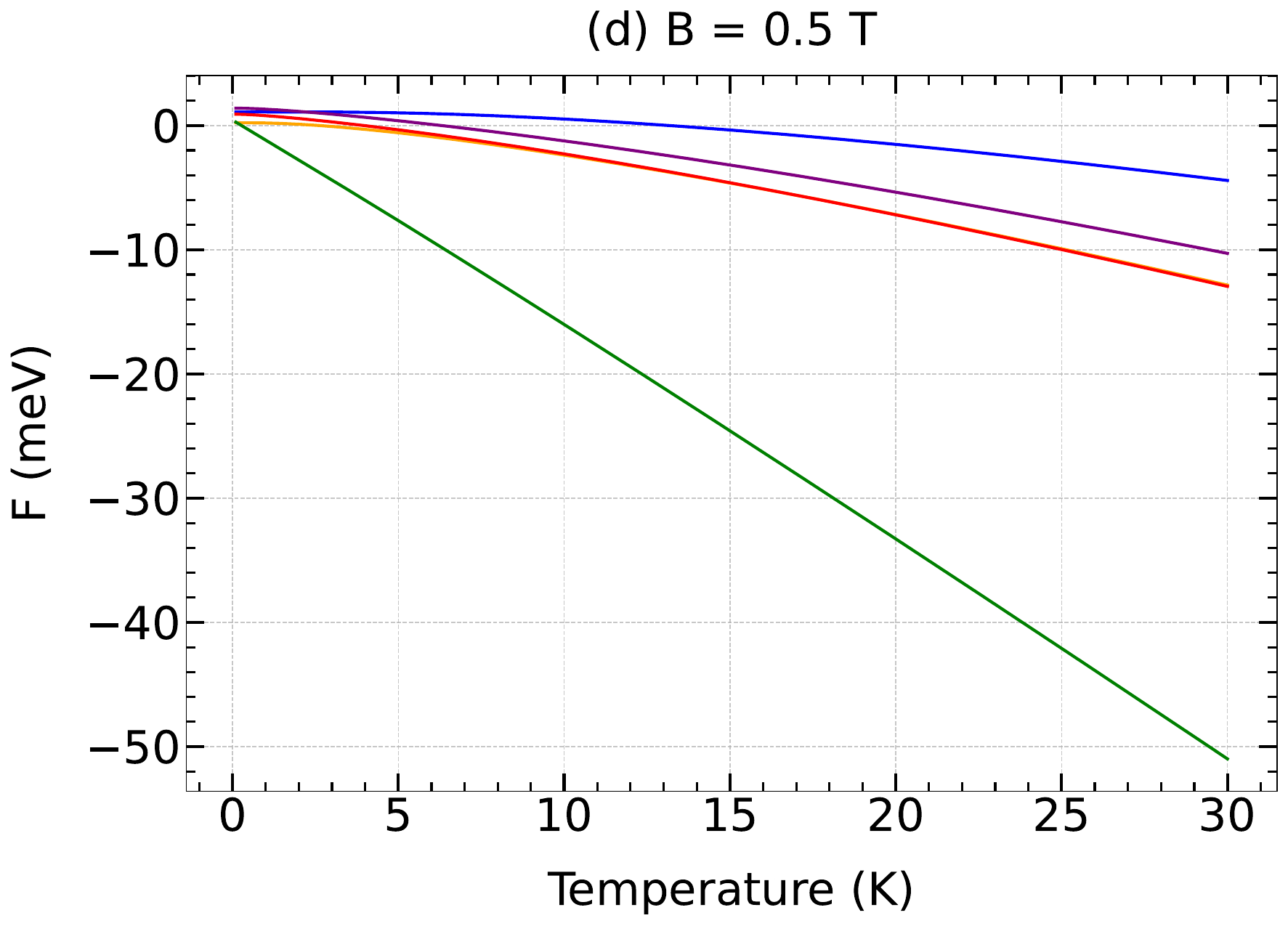}}
\caption{Free energy for a rotating 2DEG for some values of angular speed $\Omega$ as a function of temperature and for different values of an external magnetic field intensity (measured in Tesla). The plots where $\Omega = 0$ Hz were computed for the problem of degenerate Landau levels. Here, $m^*\neq m_G$.}
\label{freenergym}
\end{figure}

In Figure~\ref{freenergym}-(a), we have the $B=0$ T case. The curves corresponding to rotation frequencies of $\pm 50\,\mathrm{GHz}$ coincide with each other, as do the curves for $\pm 100\,\mathrm{GHz}$. The $\pm 50\,\mathrm{GHz}$ curves also show the greatest downward deviation.

In Figurs~\ref{freenergym}-(b), (c), and (d), the curves for $\Omega=0$ Hz are shown, representing the non-rotating case (Landau levels). At $T=0$ K, this curve shows a $F$ value lower than the rotating cases, indicating a lower ground state energy for the purely magnetic system. The presence of rotation alters the energy spectrum to raise the value of $F$. Moreover, the hierarchy among the curves changes when $B=0.5$ T, which depends on the balance between internal energy and the entropic contribution. 

\begin{figure}[t!]
{\includegraphics[width=0.48\linewidth]{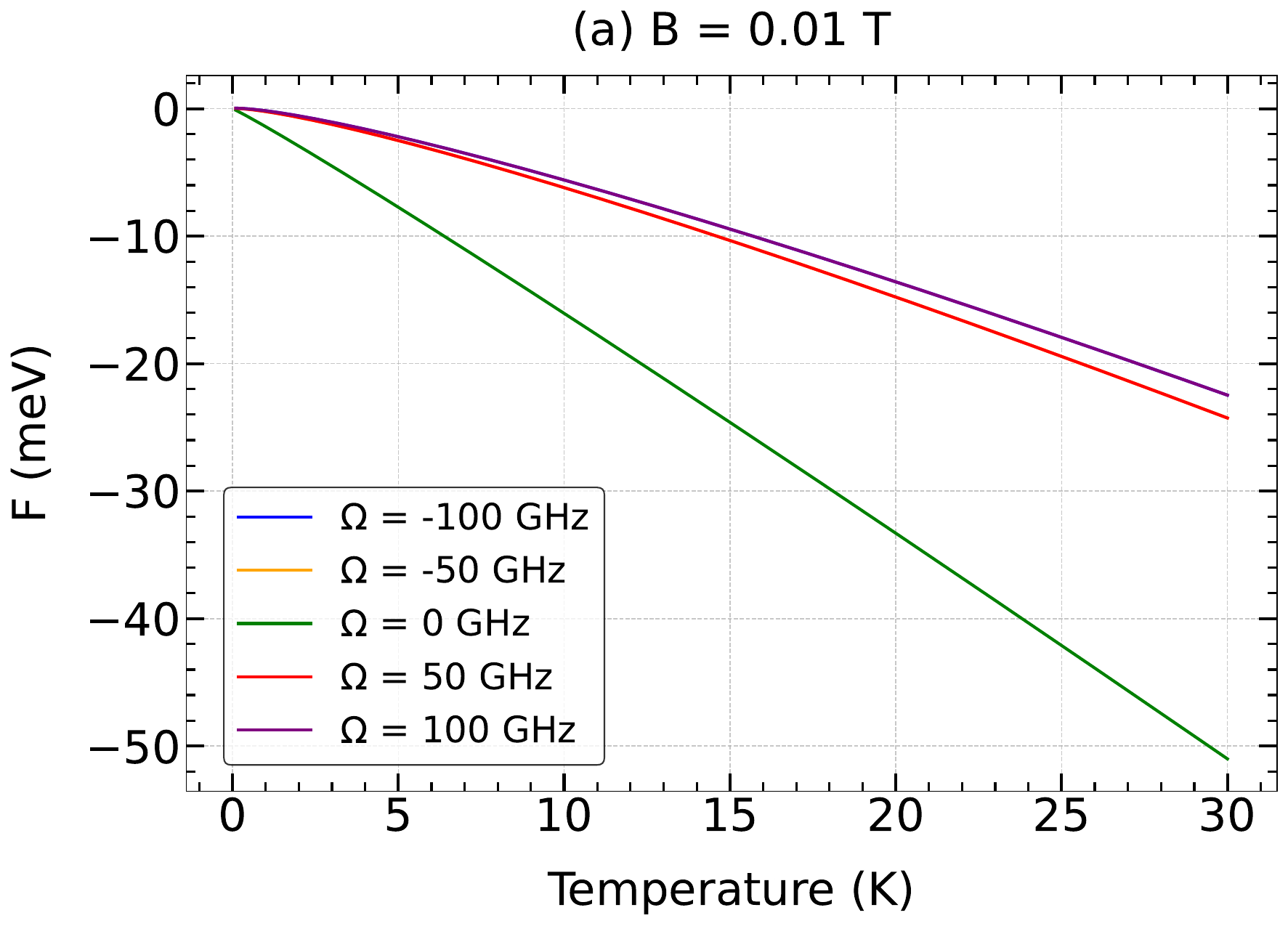}}\qquad \vspace{0.3cm}
{\includegraphics[width=0.48\linewidth]{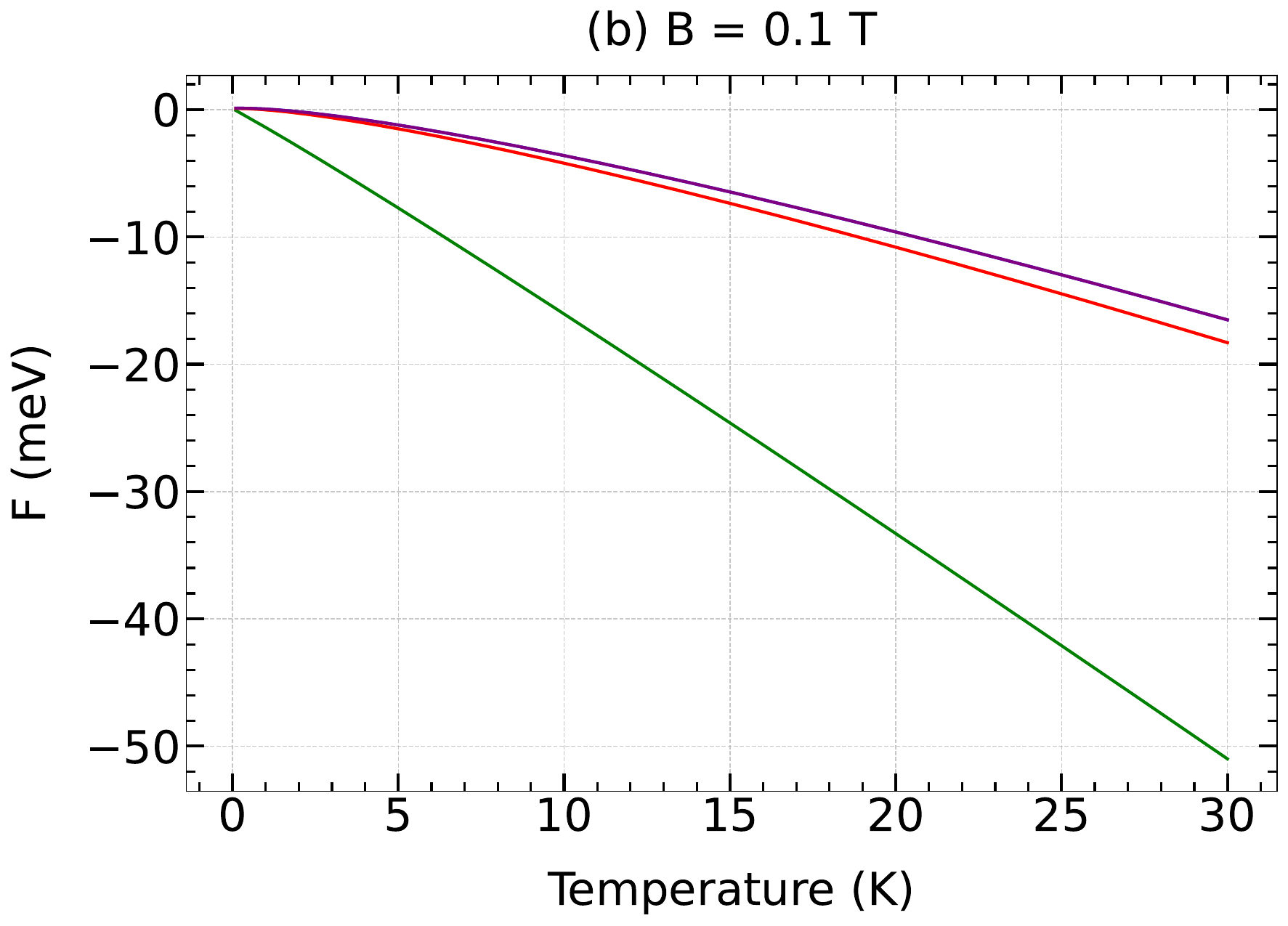}}
{\includegraphics[width=0.48\linewidth]{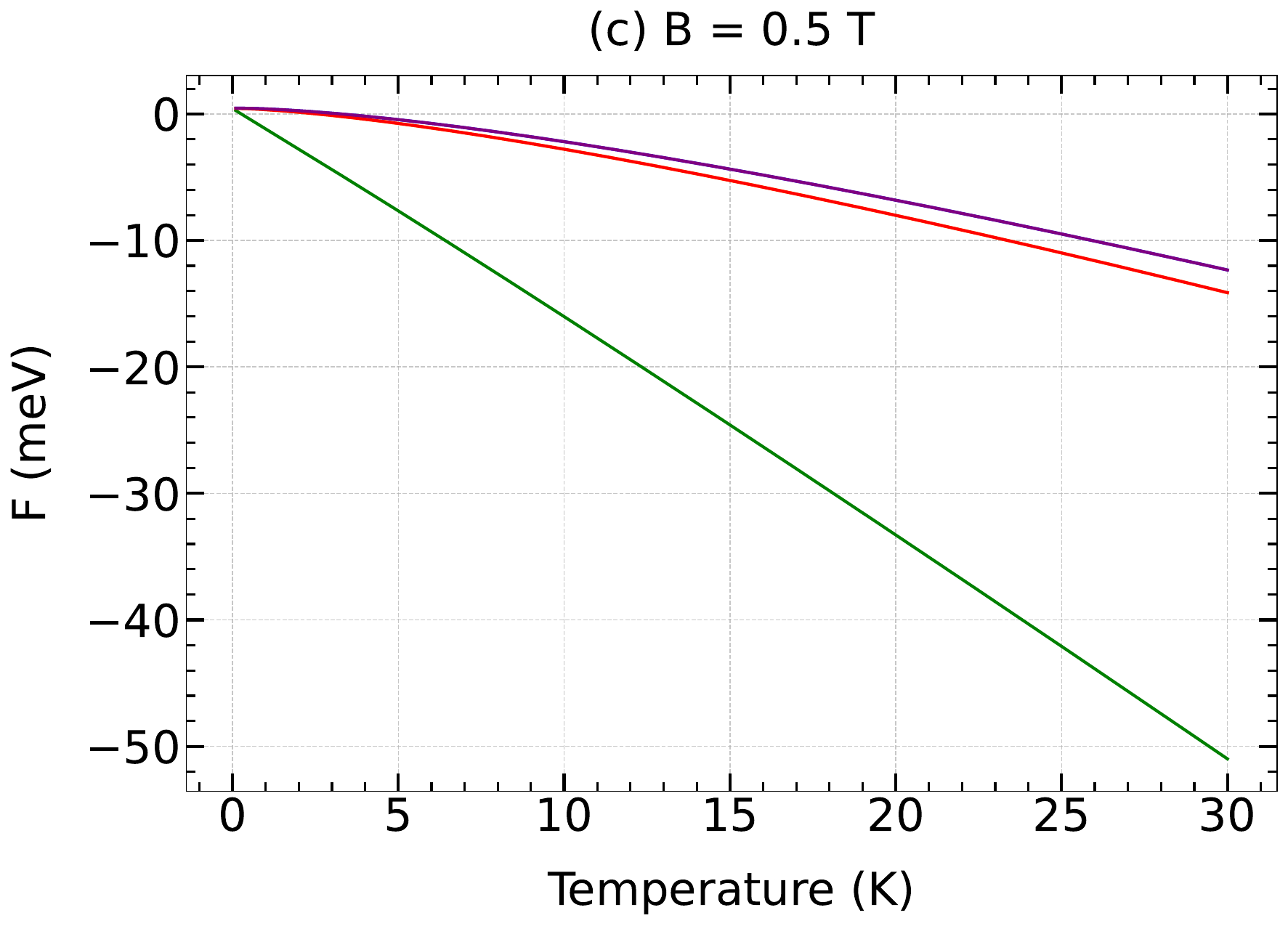}}
\caption{Free energy for a rotating 2DEG for some values of angular speed $\Omega$ as a function of temperature and for different values of an external magnetic field intensity (measured in Tesla). The plots where $\Omega = 0$ Hz were computed for the problem of degenerate Landau levels. Here, $m^*\equiv m_G$.}
\label{freenergy}
\end{figure}

Figure \ref{freenergy} considers the case where $m^* \equiv m_{G}$, showing that the rotation $\Omega$ has a significant impact on the free energy; again, noticeable deviations between curves corresponding to different rotations considered are not meaningful.

\FloatBarrier
\subsection{Entropy}

\begin{figure}[t!]
{\includegraphics[width=0.45\linewidth]{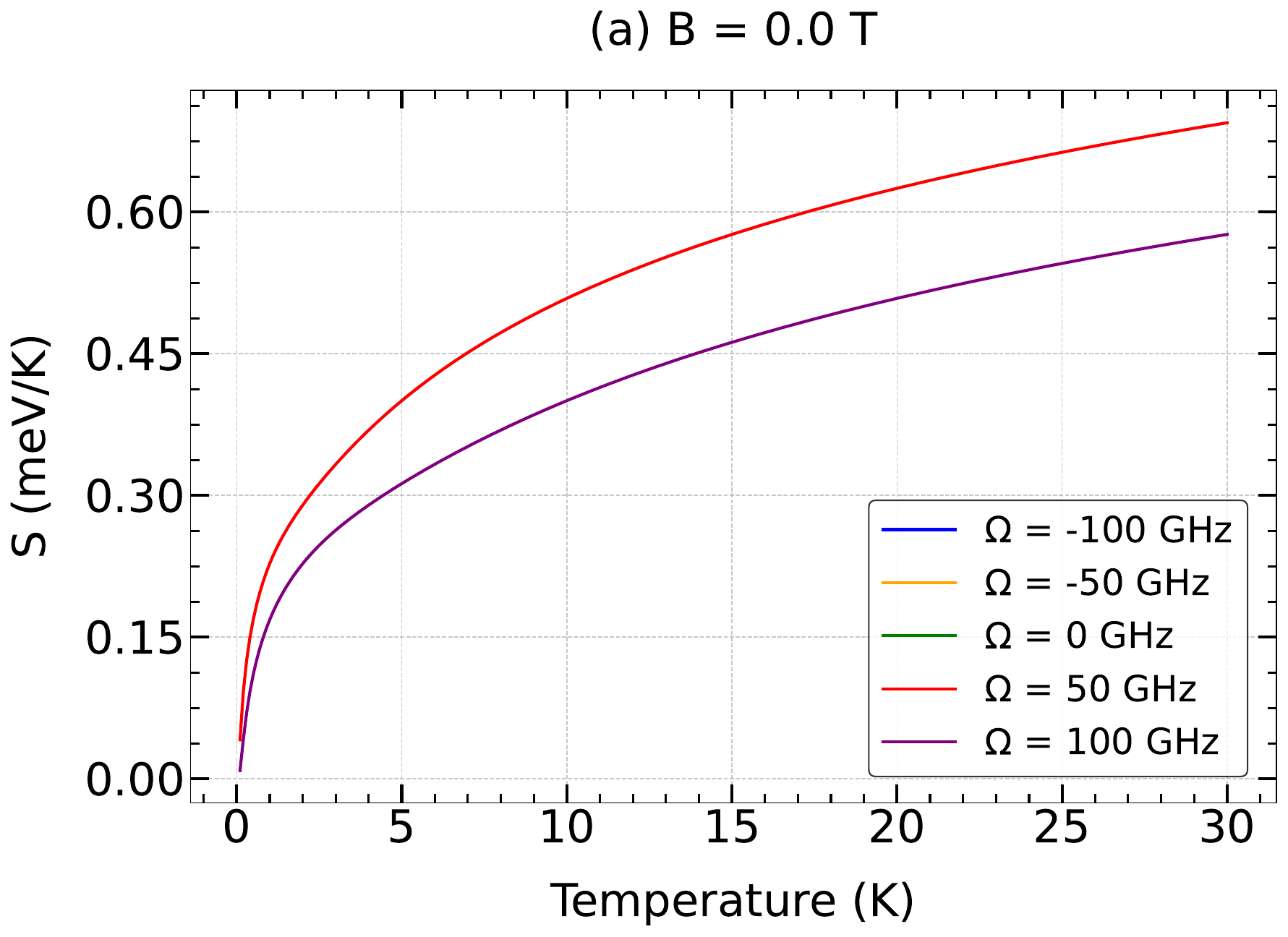}}\qquad \vspace{0.3cm}
{\includegraphics[width=0.45\linewidth]{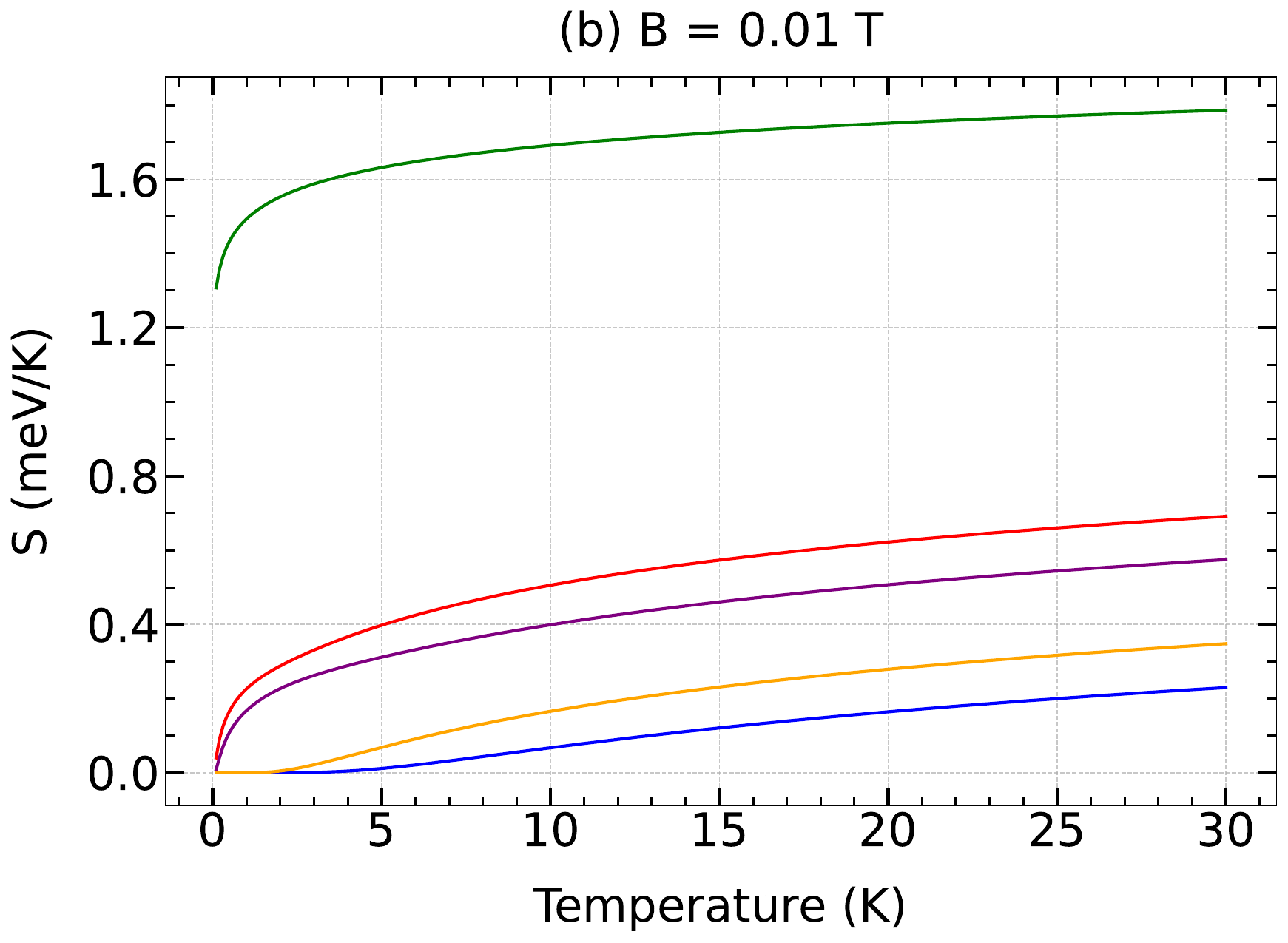}}
{\includegraphics[width=0.45\linewidth]{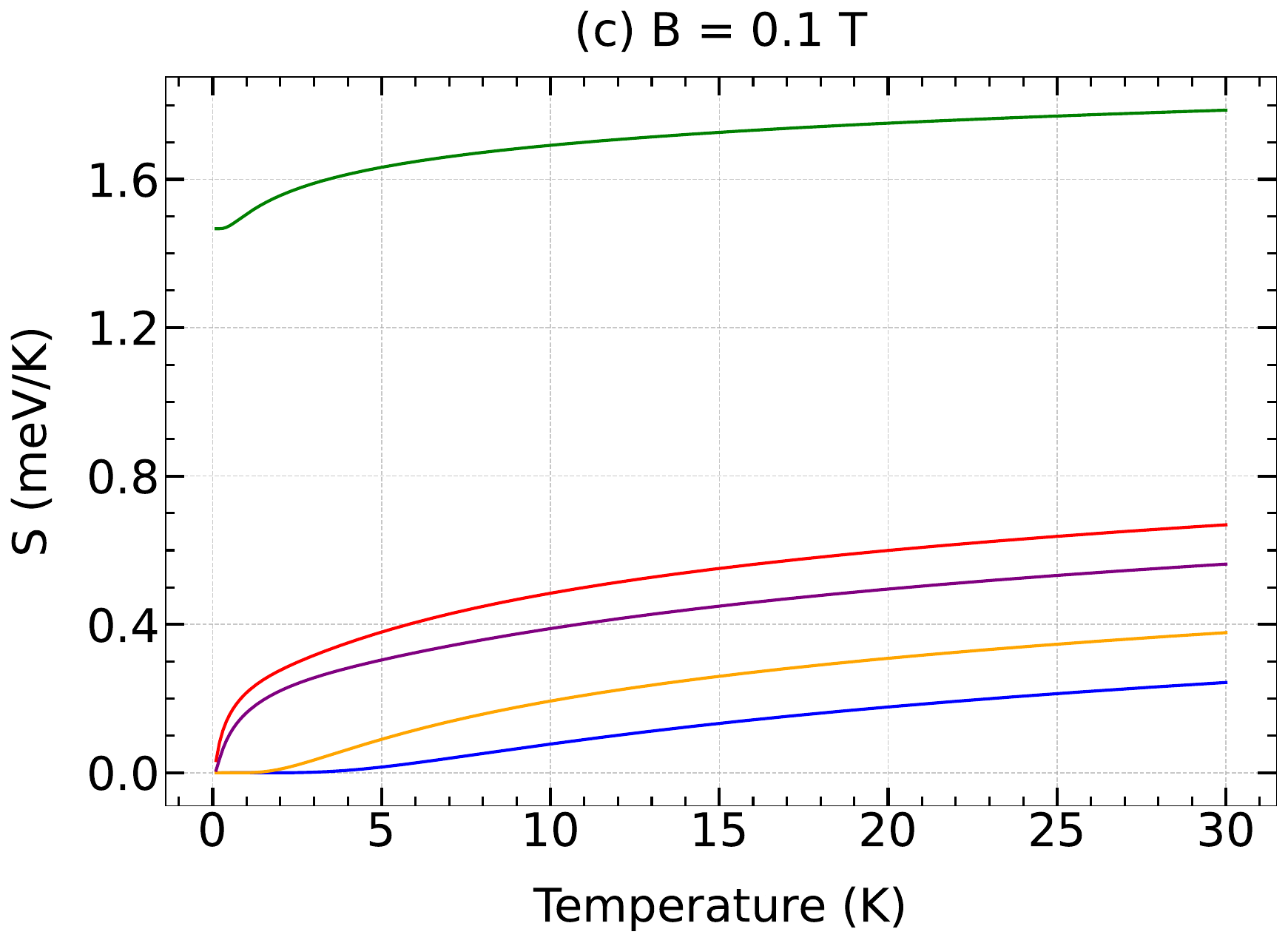}}\qquad
{\includegraphics[width=0.45\linewidth]{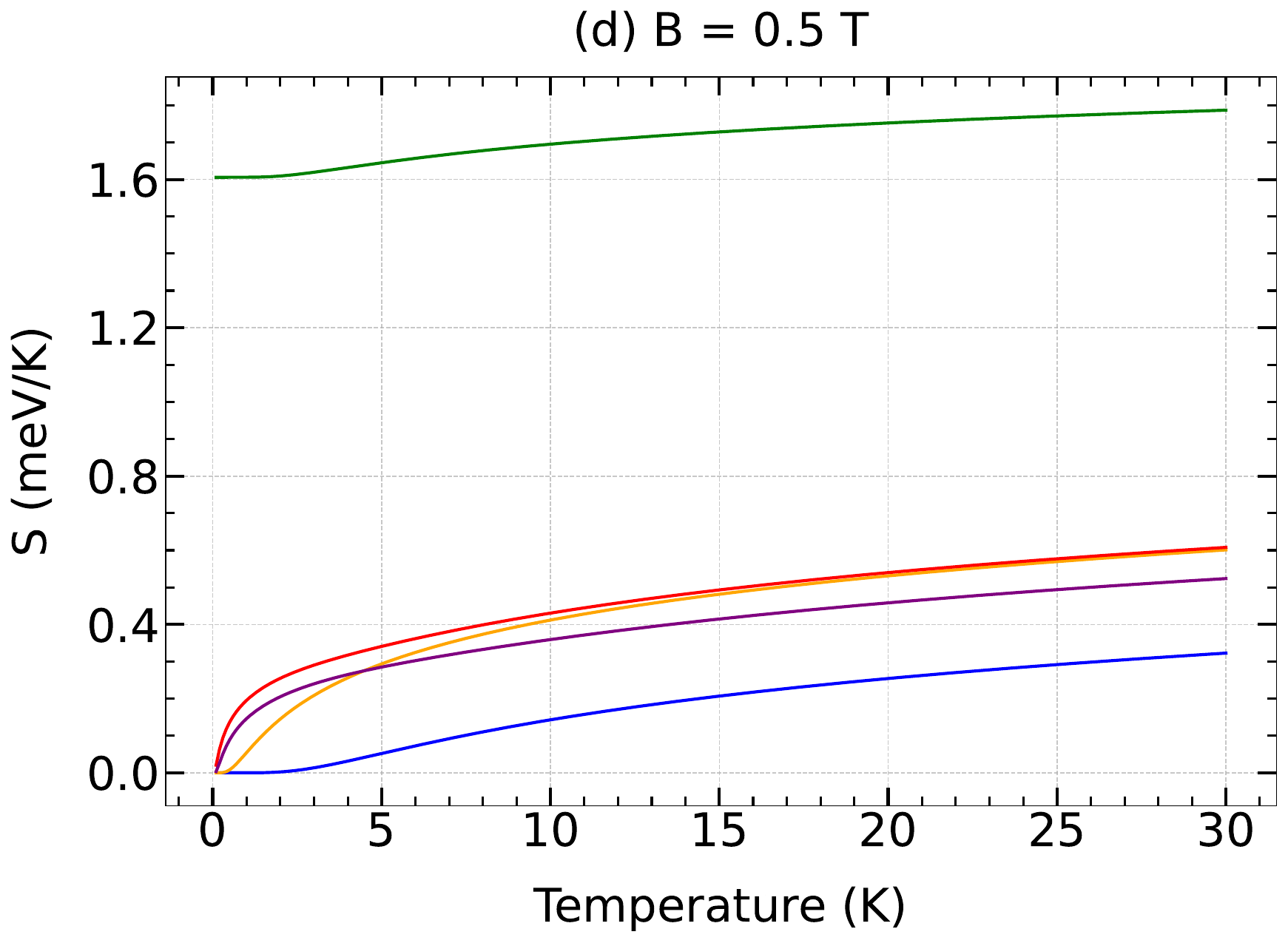}}
\caption{Entropy for a rotating 2DEG for some values of angular speed $\Omega$ as a function of temperature and for different values of an external magnetic field intensity (measured in Tesla). The plots where $\Omega = 0$ Hz were computed for the problem of degenerate Landau levels. Here, $m^*\neq m_G$.}
\label{entropym}
\end{figure}
\begin{figure}[t!]
{\includegraphics[width=0.48\linewidth]{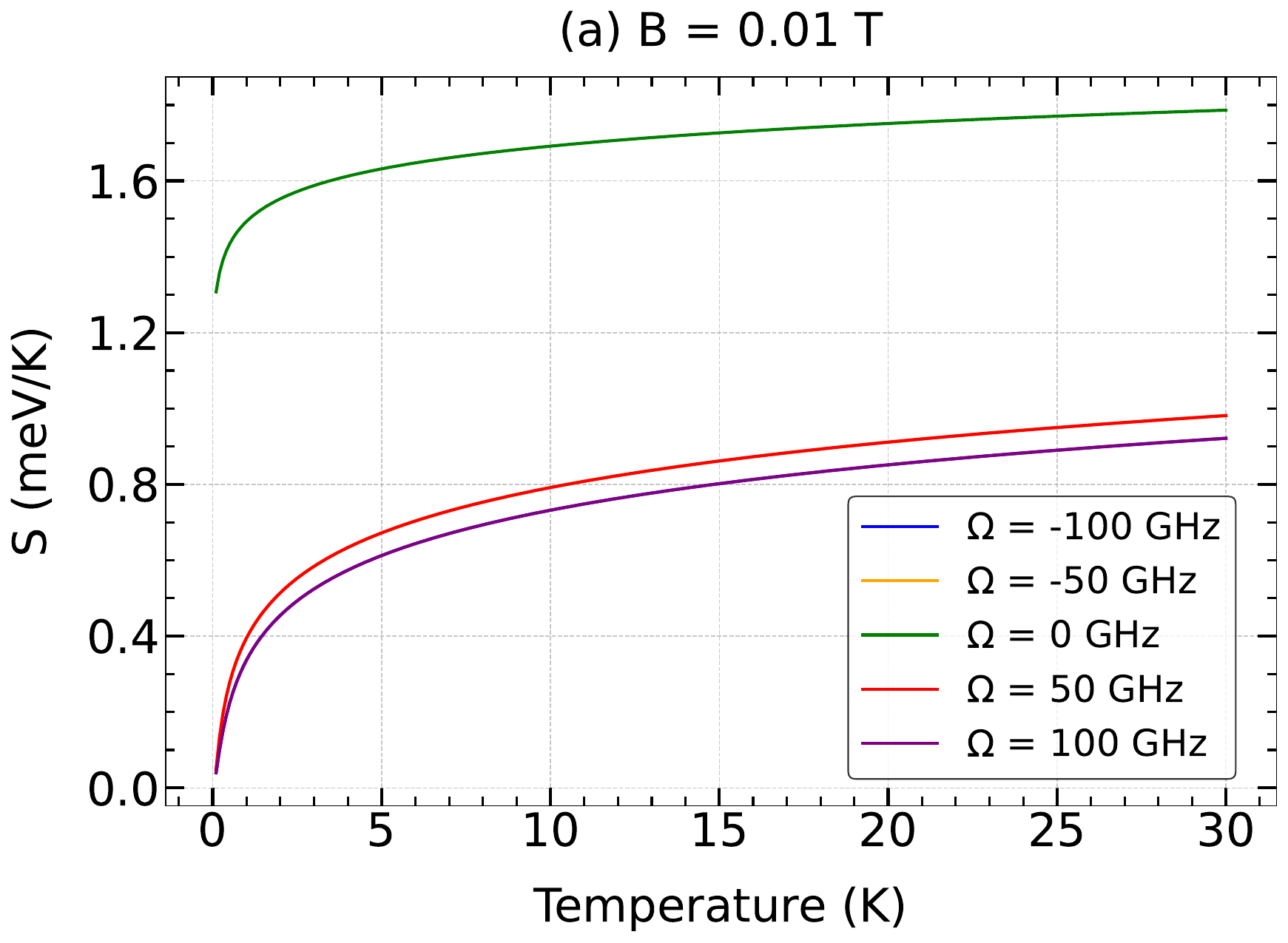}}\qquad \vspace{0.3cm}
{\includegraphics[width=0.48\linewidth]{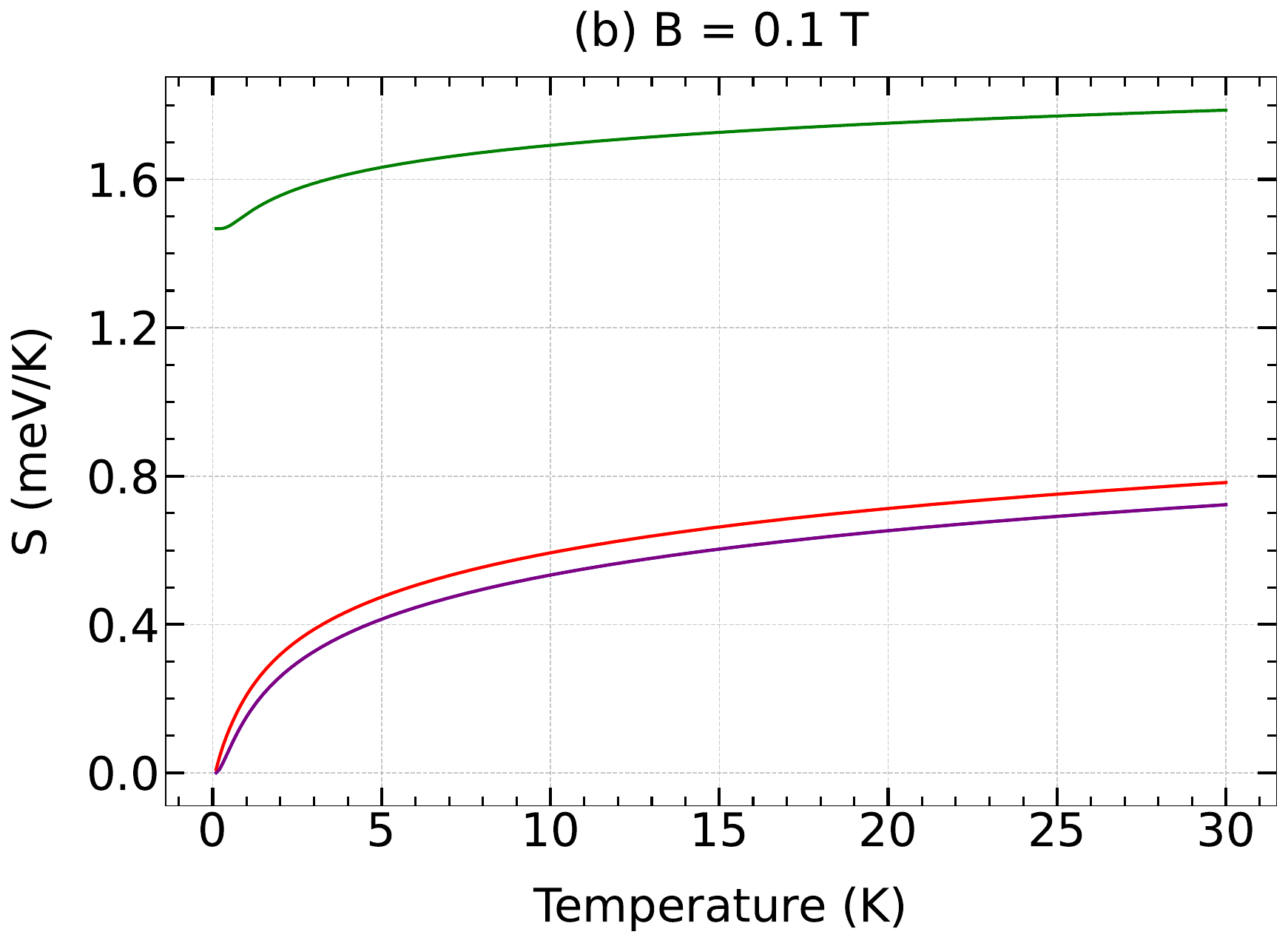}}
{\includegraphics[width=0.48\linewidth]{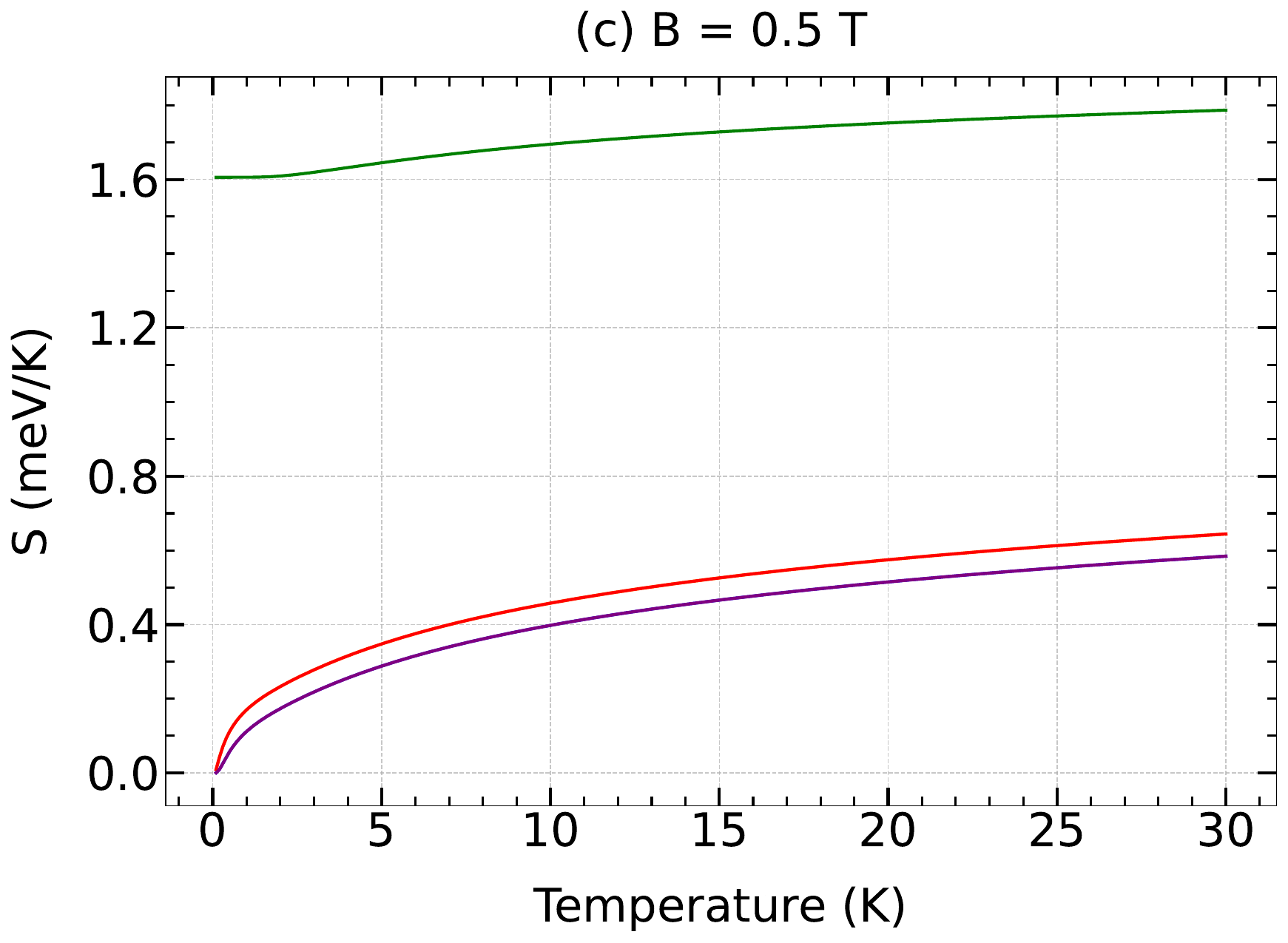}}
\caption{Entropy for a rotating 2DEG for some values of angular speed $\Omega$ as a function of temperature and for different values of an external magnetic field intensity (measured in Tesla). The plots where $\Omega = 0$ Hz were computed for the problem of degenerate Landau levels. Here, $m^*\equiv m_G$.}
\label{entropy}
\end{figure}
Figures~\ref{entropym}-(a), (b), (c), and (d) display the entropy ($S$) as a function of temperature ($T$) for some different magnetic field values for different rotation scenarios. The curves corresponding to rotation frequencies of $\pm 50\,\mathrm{GHz}$ at $B=0$ T coincide with each other, as do the curves for $\pm 100\,\mathrm{GHz}$. Additionally, the $\pm 50\,\mathrm{GHz}$ curves show higher values. The $B \neq 0$ T curve corresponds to the non-rotating case, where Landau levels exhibit high ground state degeneracy, resulting in a nonzero entropy value even at $T=0$ K. All curves start from $S=0$ eV/K at $T=0$ K, in accordance with the Third Law of Thermodynamics for systems with a non-degenerate ground state. As the temperature increases, $S$ grows, reflecting the increasing number of accessible excited states. The ($+50\,\mathrm{GHz}$) curves reach higher values than the others throughout the $T$ range. There is a change in the hierarchy between the curves at $-50$ GHz compared to the other two for $B=0.5$ T. Although rotation modifies the density of states, the effect of the intense magnetic field dominates the entropy behavior, keeping it very high for the non-rotating case. The combination of magnetic field and rotation affects the energy level distribution and, consequently, the system's entropy in a significant manner.

Figure \ref{entropy} considers the case where $m^* \equiv m_{G}$, showing that the rotation $\Omega$ still has a significant impact on the entropy; however, noticeable deviations between curves corresponding to different rotations considered are not observed.

The results demonstrate a consistent interplay among the studied thermodynamic properties. The presence of the magnetic field enhances the internal energy through Landau levels, leading to lower free energy and higher entropy, even at low temperatures. Simultaneously, rotation modifies the energy level spacing, influencing the specific heat and internal energy in a nontrivial way, evidenced by the inversions in curve hierarchy as temperature varies. Together, these effects reveal that the magnetic field and rotation act complementarily in defining the system's overall behavior.

\FloatBarrier
\subsection{Magnetization}

\begin{figure}[hbt!] {\includegraphics[width=0.46\linewidth]{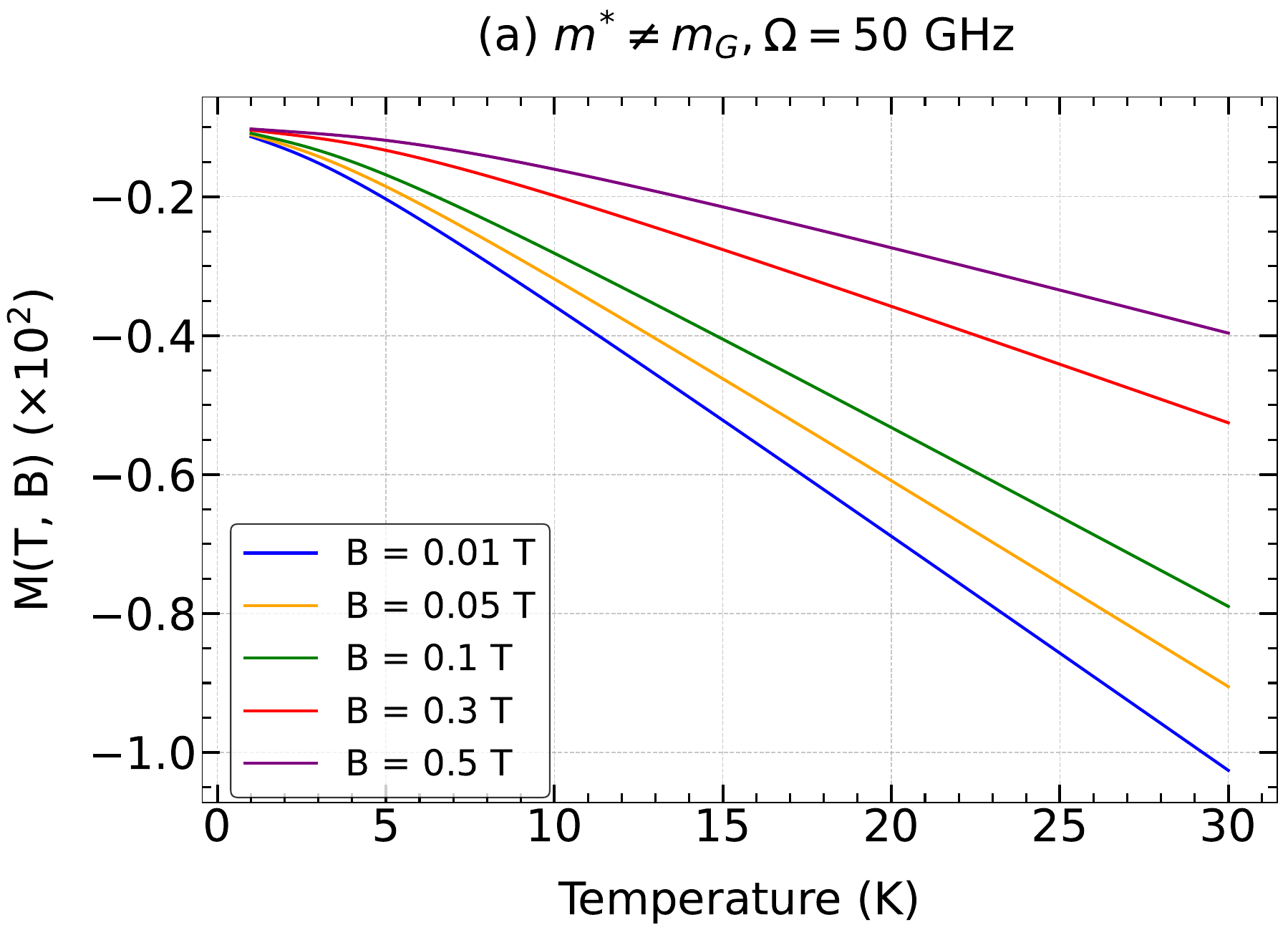}}\qquad
{\includegraphics[width=0.46\linewidth]{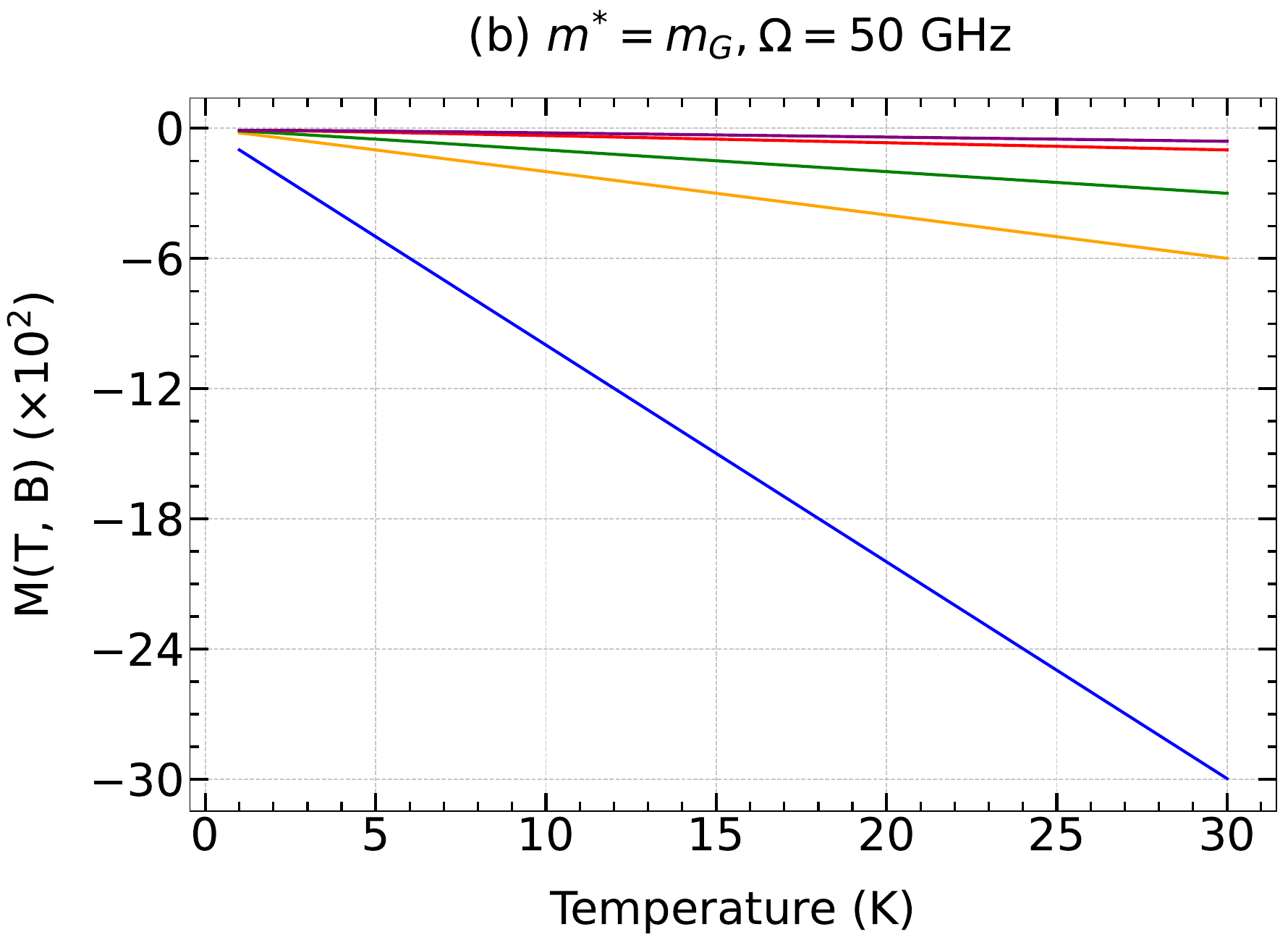}} 
\caption{Magnetization $M(T,B)$ for a rotating 2DEG as a function of temperature, with $\Omega=50$ GHz. The main figure
shows $M(T,B)$ for some values of external magnetic fields in Teslas, up to 30 K in the temperature scale, when (a) $m^* \neq m_G$ and (b) $m^* \equiv m_G$.}
\label{mag50}
\end{figure}

\begin{figure}[hbt!]{\includegraphics[width=0.46\linewidth]{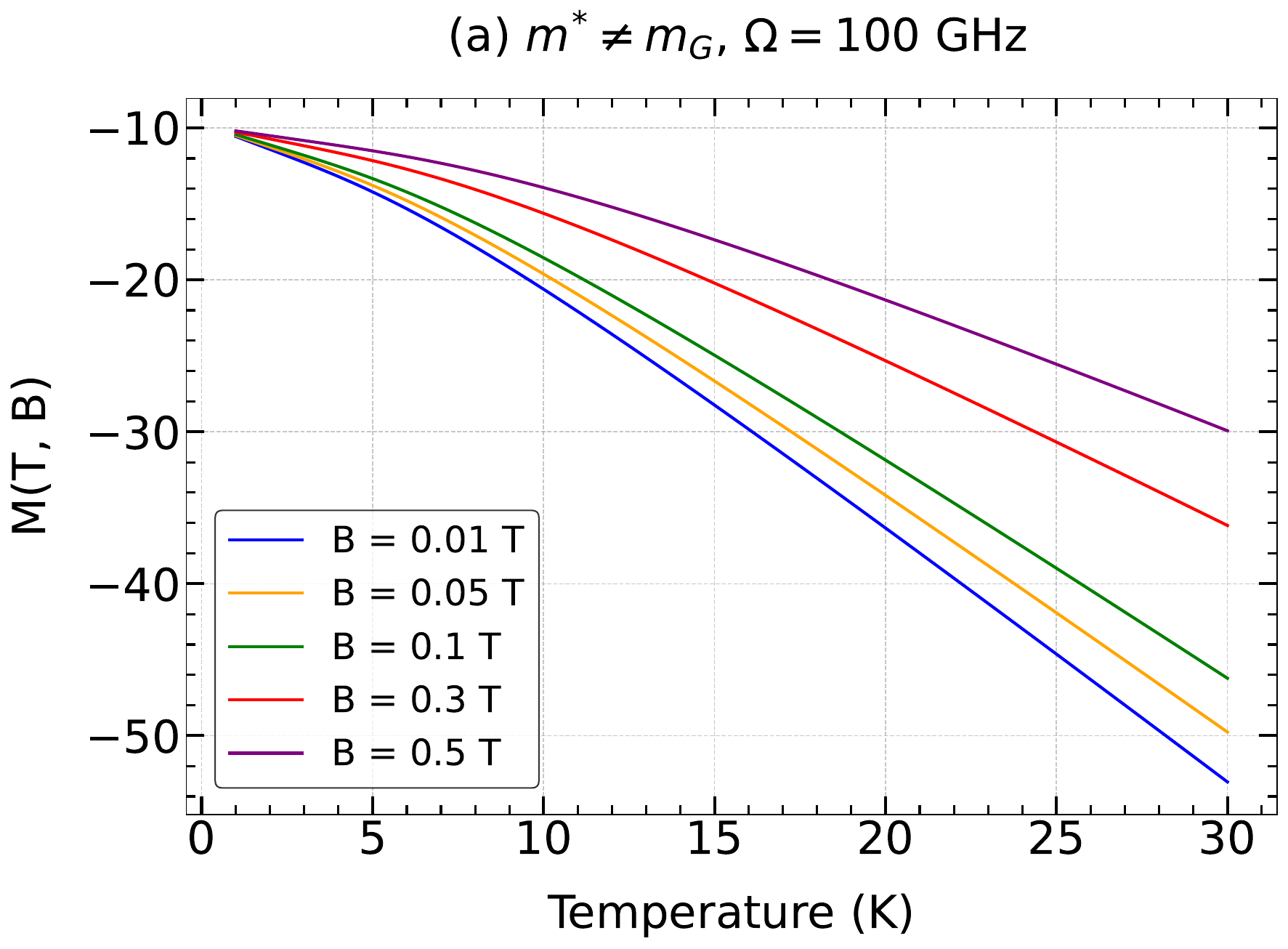}}\qquad
{\includegraphics[width=0.46\linewidth]{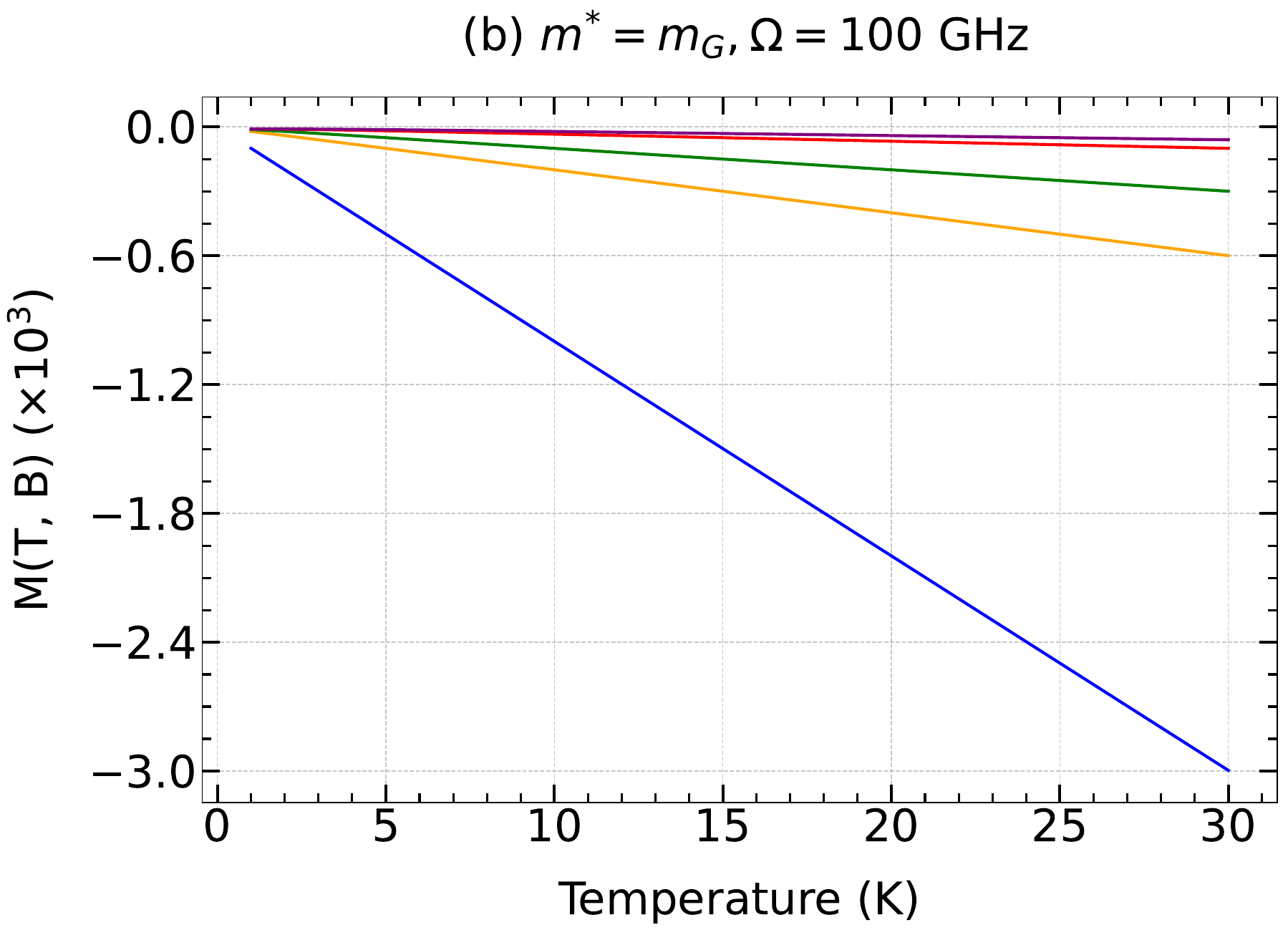}}
\caption{Magnetization $M(T,B)$ for a rotating 2DEG as a function of temperature, with $\Omega=100$ GHz. The main figure shows $M(T,B)$ for some values of external magnetic fields in Teslas, up to 30 K in the temperature scale, when a) $m^* \neq m_G$ and b) $m^* \equiv m_G$.}
\label{mag100}
\end{figure}

In Figs.~\ref{mag50} and \ref{mag100}, the magnetization $M(T, B)$ decreases (becomes more negative) with increasing temperature for all values of $B$, indicating a typical thermal response of magnetic systems, where thermal agitation tends to reduce magnetic ordering. It is observed that the magnetization exhibits a higher magnitude when $m^* \equiv m_{G}$.

\begin{figure}[hbt!] {\includegraphics[width=0.46\linewidth]{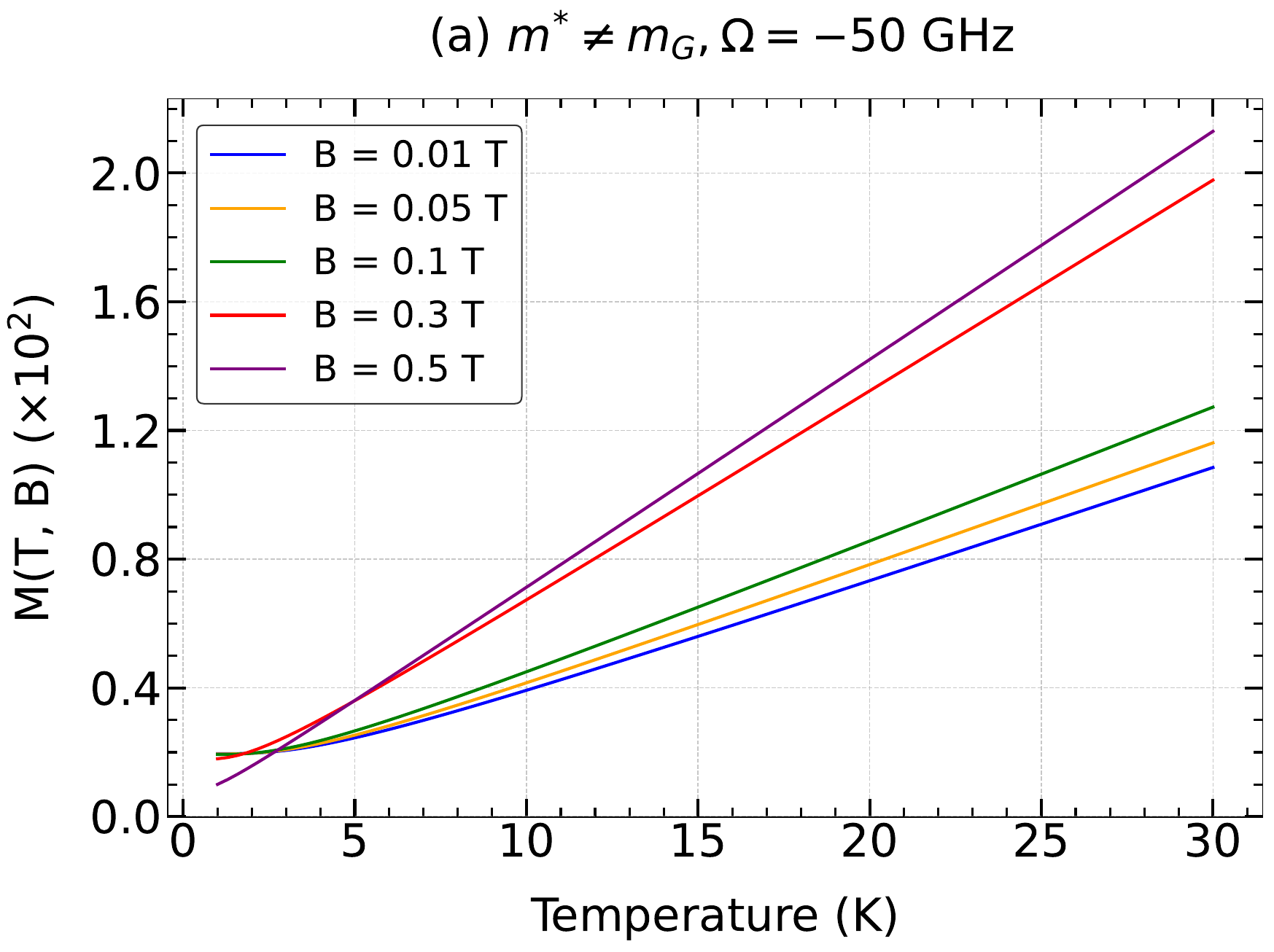}}\qquad
{\includegraphics[width=0.46\linewidth]{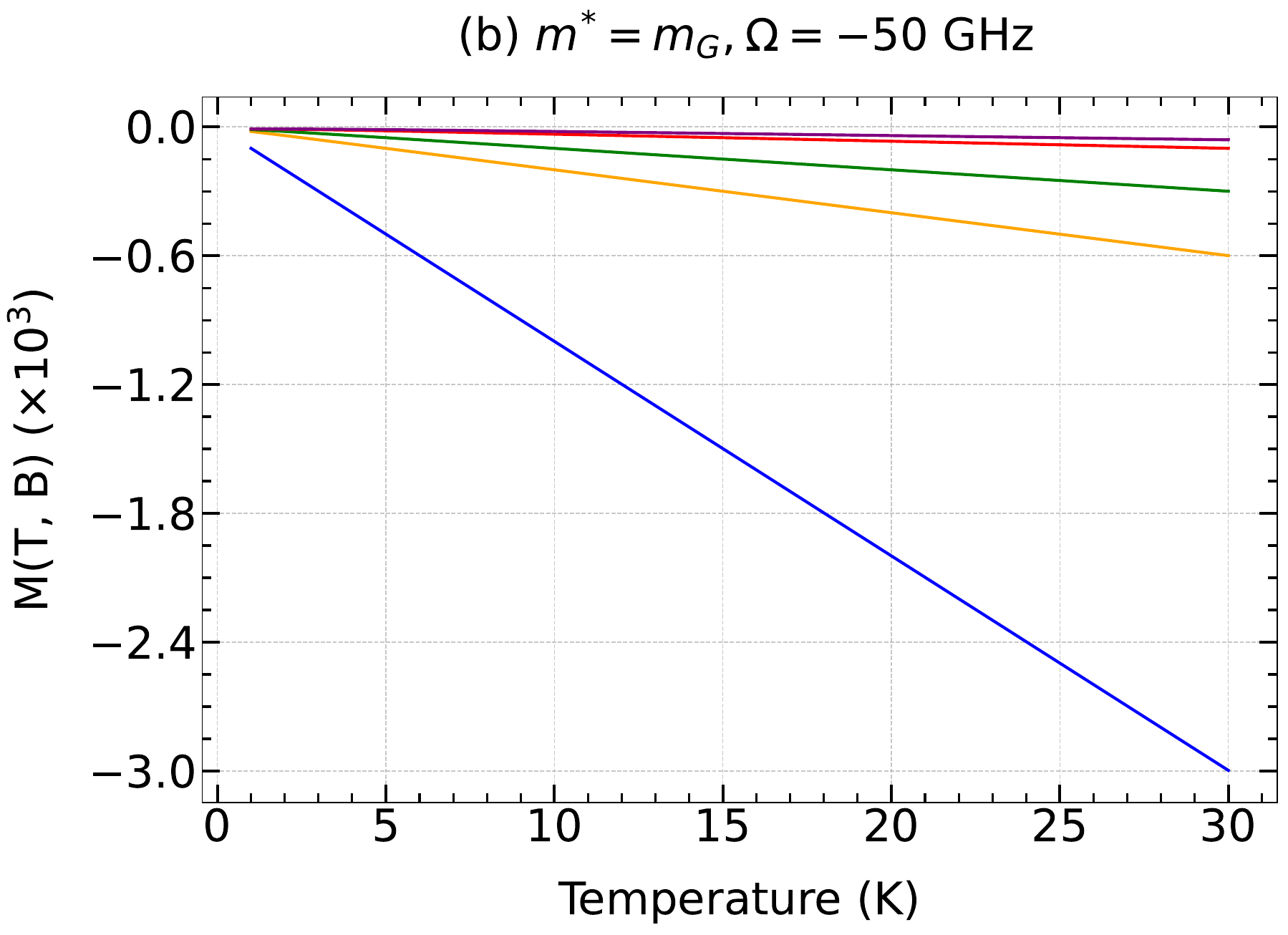}}
\caption{Magnetization $M(T,B)$ for a rotating 2DEG as a function of temperature, with $\Omega=-50$ GHz. The main figure
shows $M(T,B)$ for some values of external magnetic fields in Teslas, up to 30 K in the temperature scale, when (a) $m^* \neq m_G$ and (b) $m^* \equiv m_G$.}
\label{magminus50}
\end{figure}

\begin{figure}[hbt!]
{\includegraphics[width=0.46\linewidth]{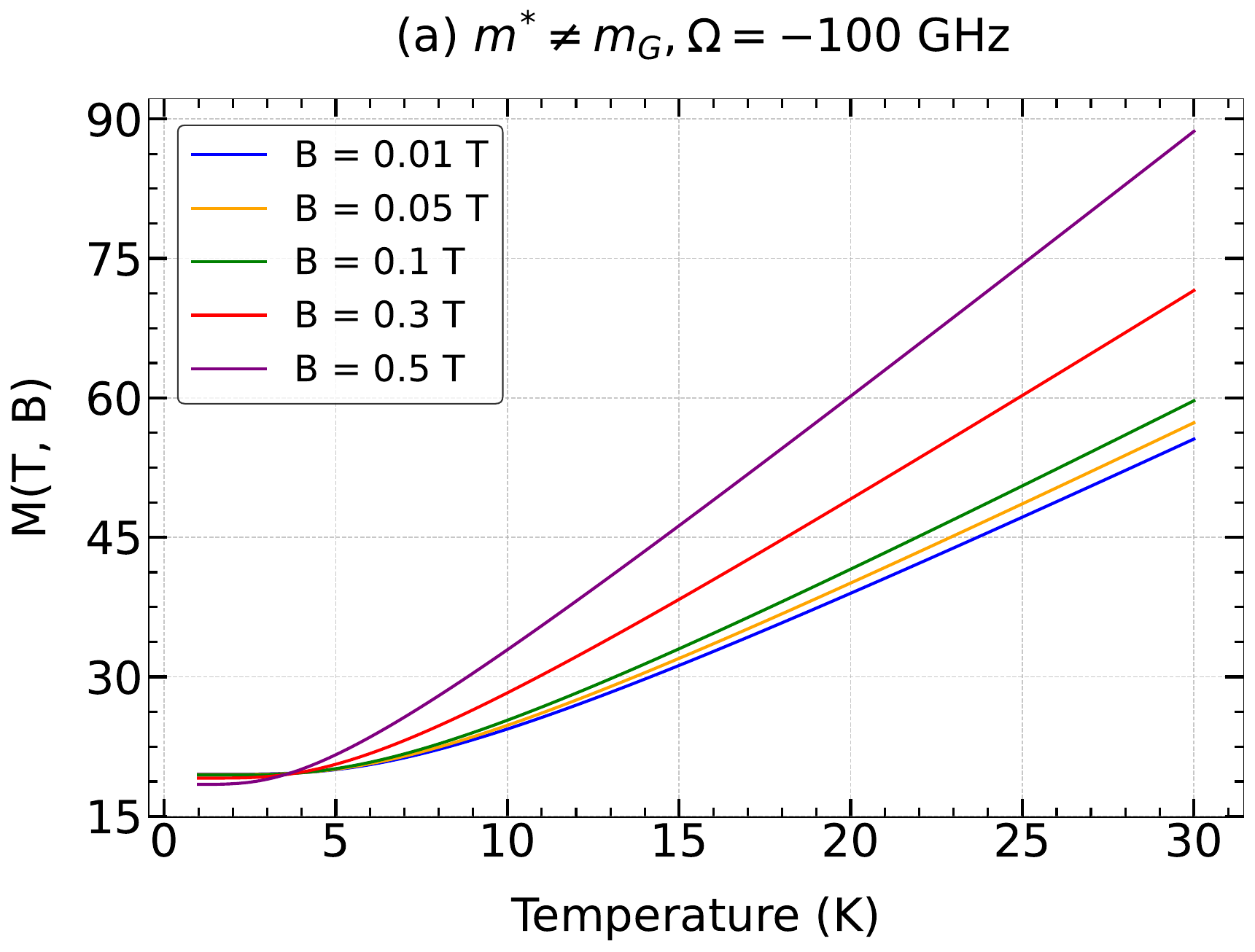}}\qquad
{\includegraphics[width=0.46\linewidth]{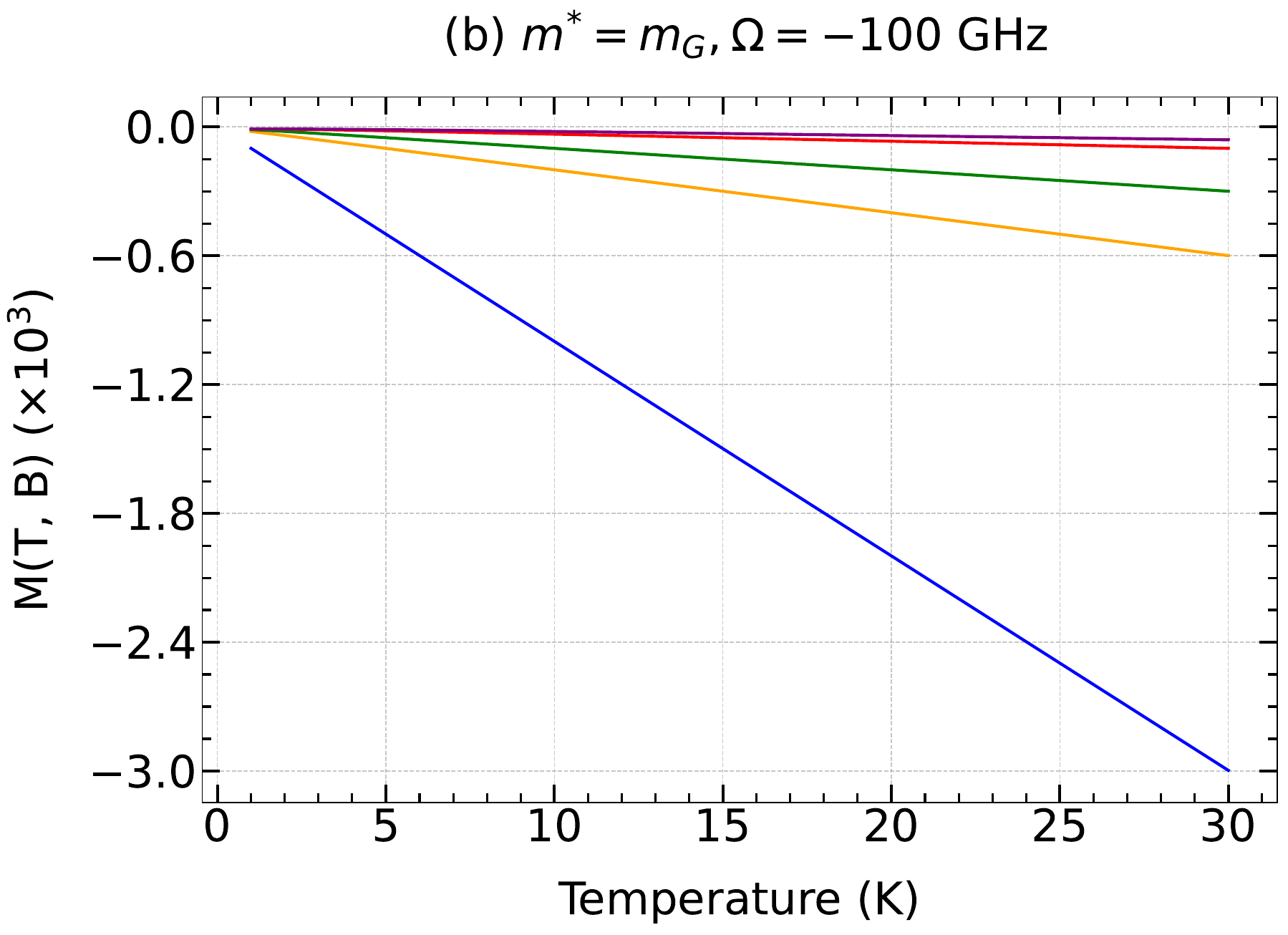}}
\caption{Magnetization $M(T,B)$ for a rotating 2DEG as a function of temperature, with $\Omega=-100$ GHz. The main figure
shows $M(T,B)$ for some values of external magnetic fields in Teslas, up to 30 K in the temperature scale, when (a) $m^* \neq m_G$ and (b) $m^* \equiv m_G$.}
\label{magminus100}
\end{figure}

Figures \ref{magminus50} and \ref{magminus100}, for $ \Omega = -50 $ GHz and $ \Omega = -100 $ GHz, respectively, it can be seen that the magnetization changes sign in the case where $ m^* \neq m_{G}$, indicating that the magnetocaloric effect, which will be investigated next, will change significantly in these cases.
\FloatBarrier

\subsection{The magnetocaloric effect}

\begin{figure}[hbt!]
{\includegraphics[width=0.48\linewidth]{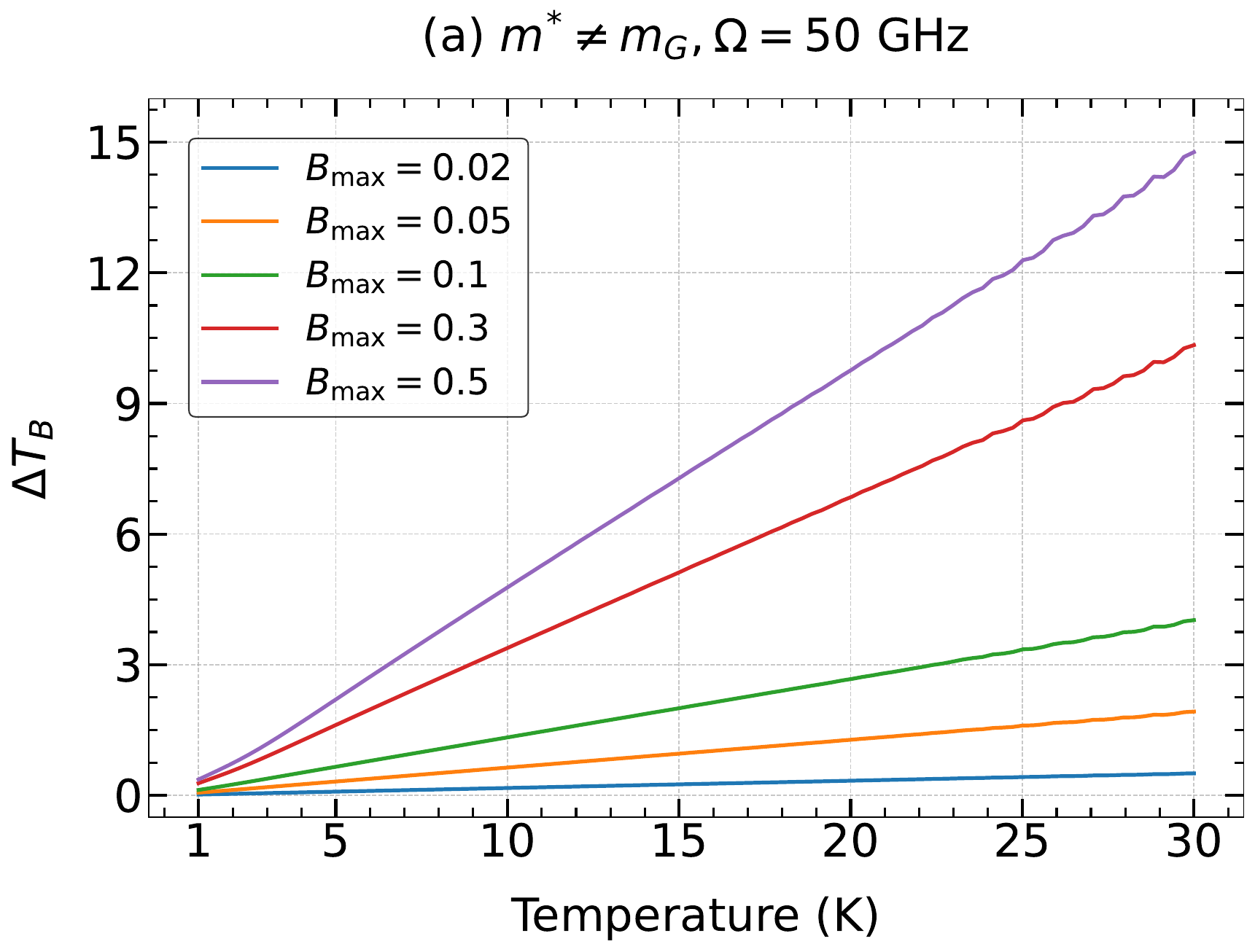}}\qquad
{\includegraphics[width=0.48\linewidth]{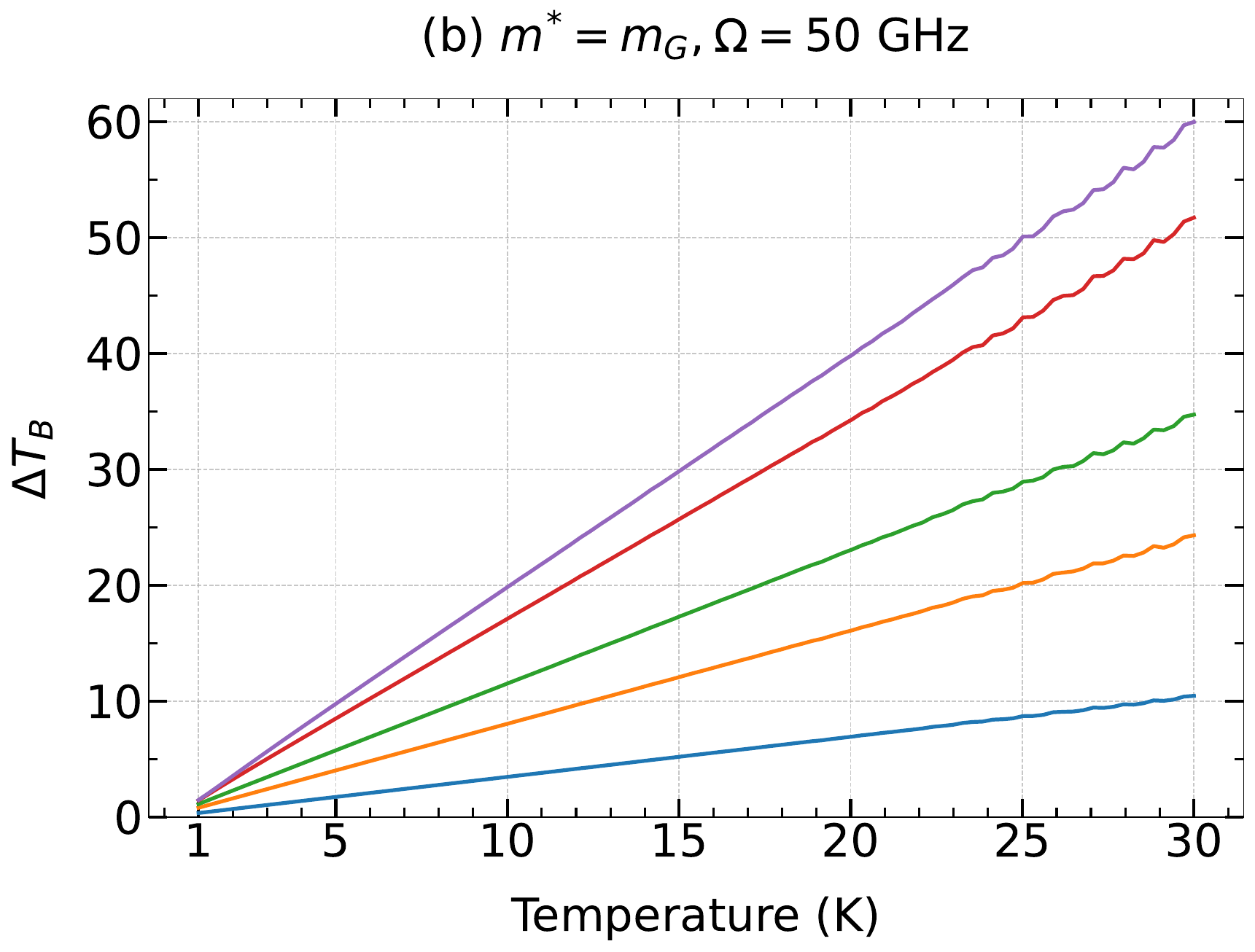}}
\caption{MCE for a rotating 2DEG as a function of temperature, with $\Omega=50$ GHz. The figures show $\Delta T_B$ for external magnetic fields $B_{\text{max}}$ at a fixed $B_i = 0.01$ T, up to 30 K in the temperature scale in the two scenarios.}
\label{mce50}
\end{figure}
\begin{figure}[hbt!]
{\includegraphics[width=0.48\linewidth]{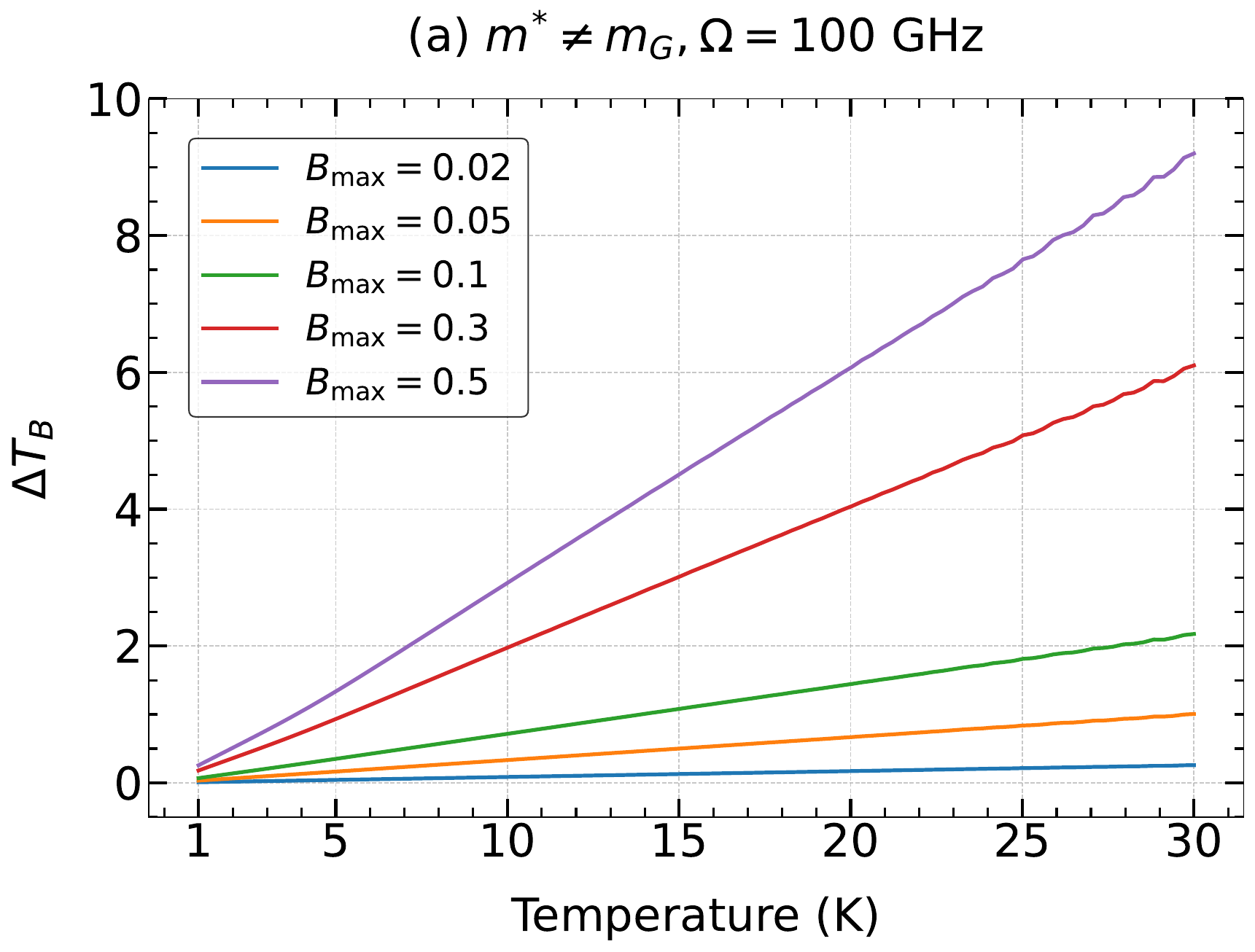}}\qquad
{\includegraphics[width=0.48\linewidth]{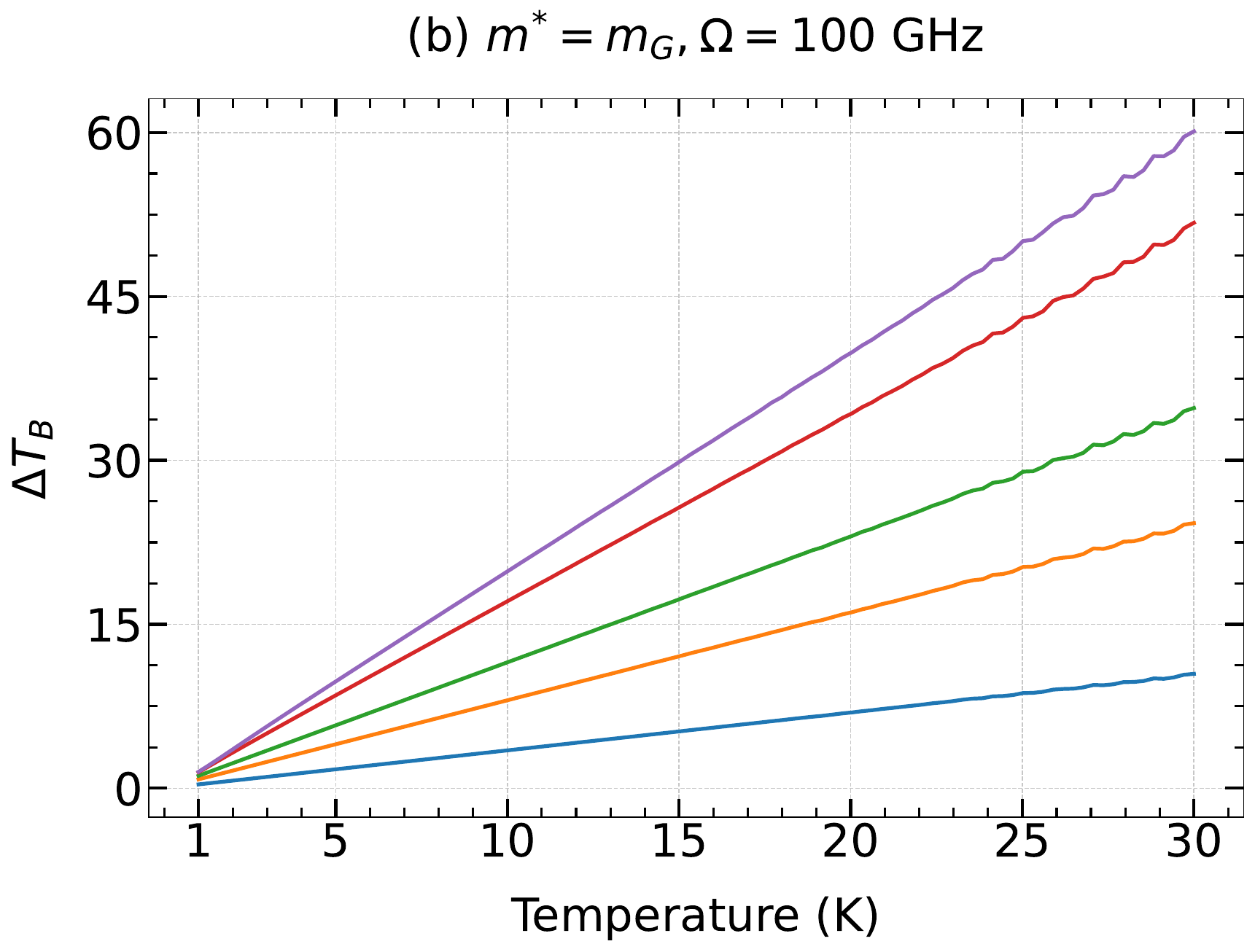}}
\caption{MCE for a rotating 2DEG as a function of temperature, with $\Omega=100$ GHz. The figures show $\Delta T_B$ for external magnetic fields $B_{\text{max}}$ at a fixed $B_i = 0.01$ T, up to 30 K in the temperature scale in the two scenarios.}
\label{mce100}
\end{figure}
\begin{figure}[hbt!]
{\includegraphics[width=0.48\linewidth]{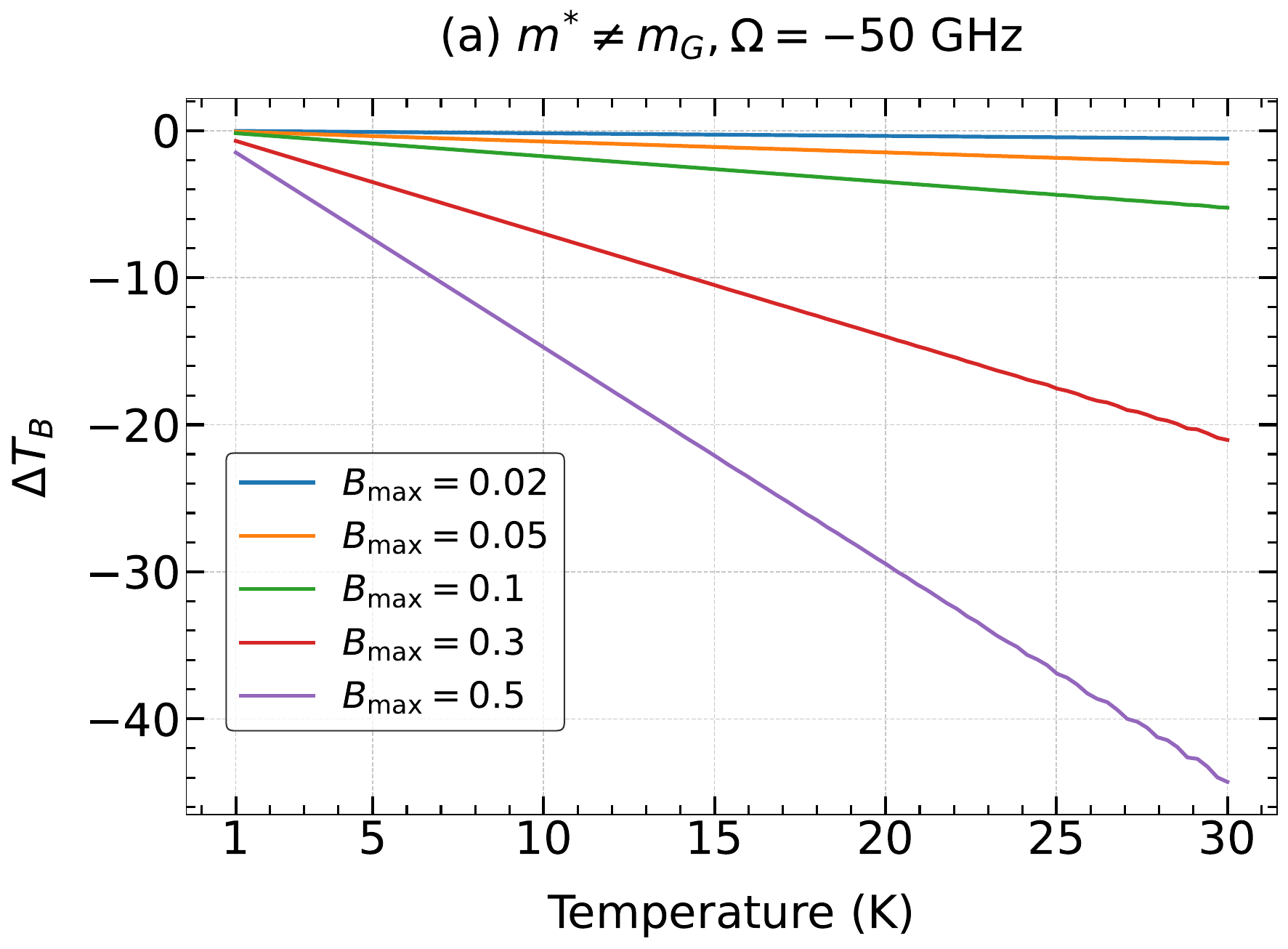}}
{\includegraphics[width=0.48\linewidth]{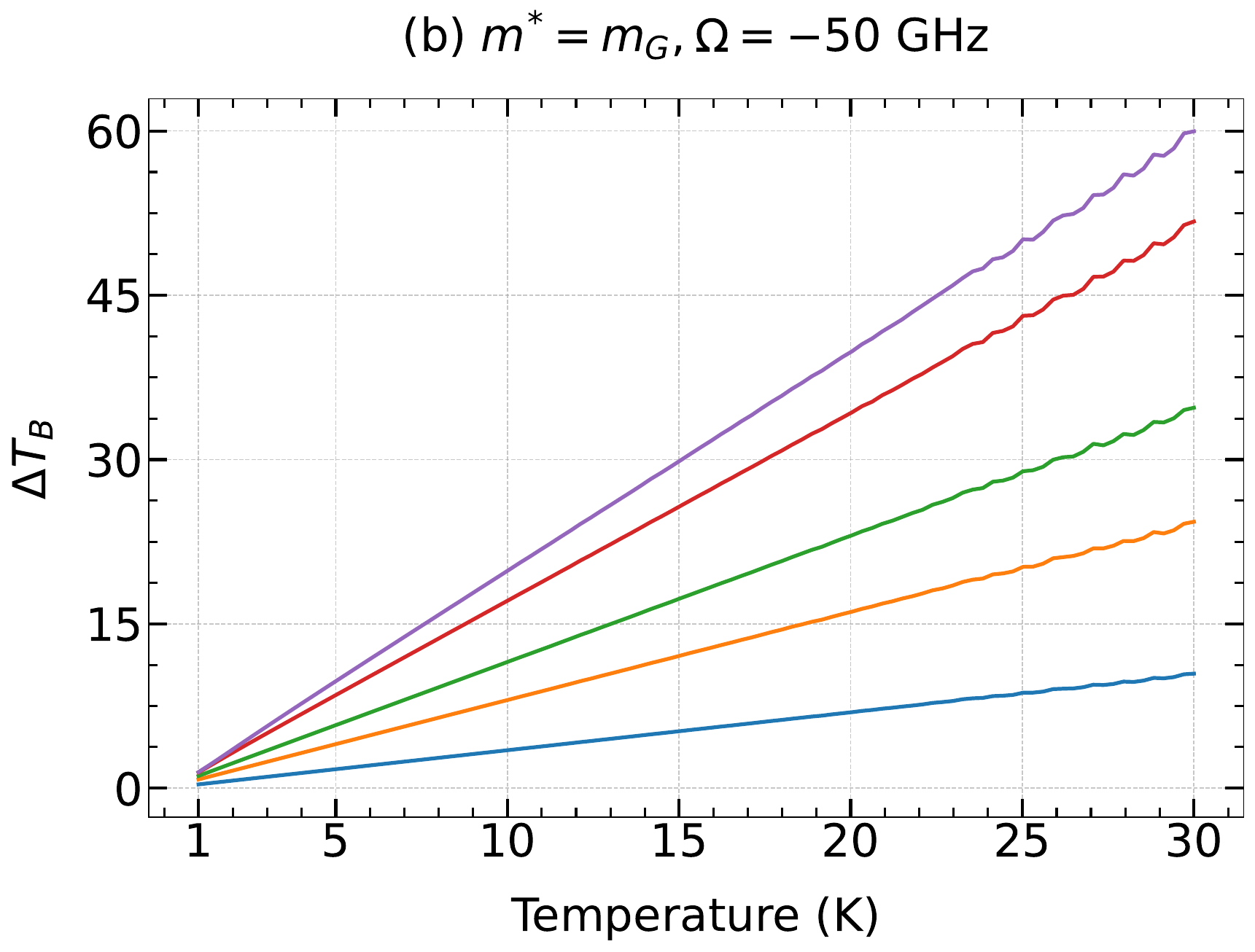}}
\caption{MCE for a rotating 2DEG as a function of temperature, with $\Omega=-50$ GHz. The figure show $\Delta T_B$ for external magnetic fields $B_{\text{max}}$ at a fixed $B_i = 0.01$ T, up to $30$ K in the temperature scale in the two scenarios. Note that cooling or heating can be predicted depending on the model considered.}
\label{mceminus50}
\end{figure}
\begin{figure}[hbt!]{\includegraphics[width=0.48\linewidth]{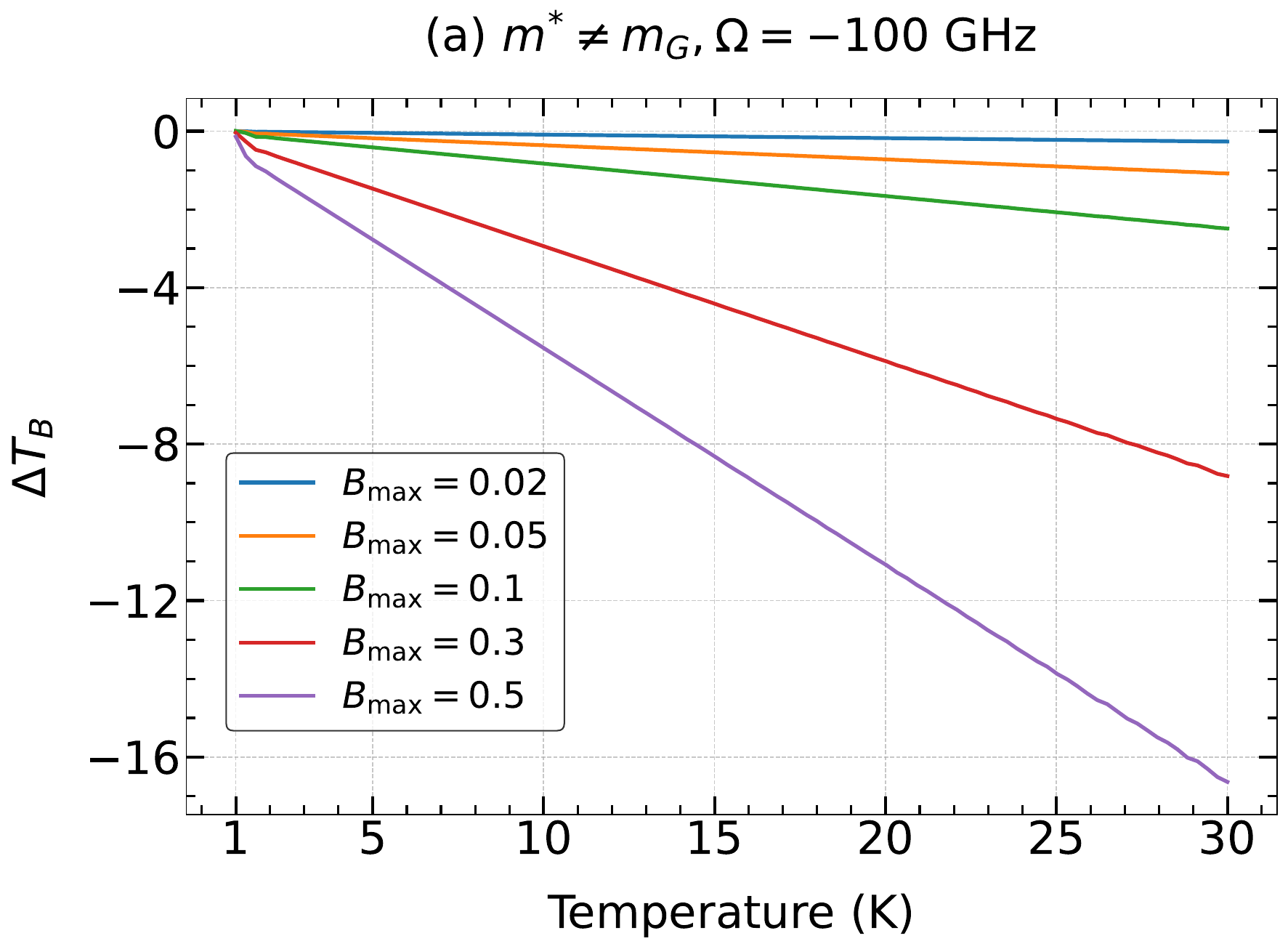}}\qquad
{\includegraphics[width=0.48\linewidth]{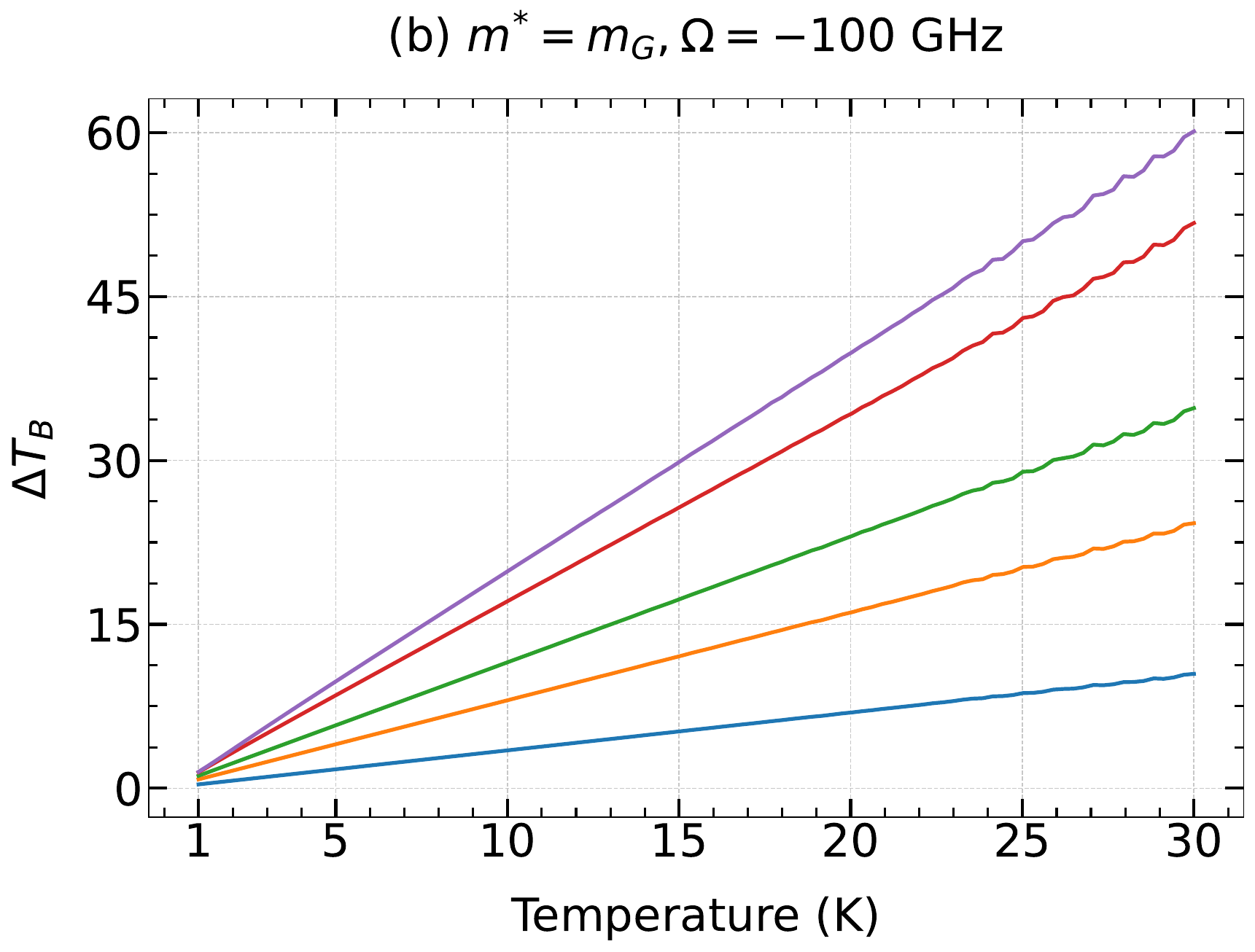}}
\caption{MCE for a rotating 2DEG as a function of temperature, with $\Omega=- 100$ GHz. The figures show $\Delta T_B$ for external magnetic fields $B_{\text{max}}$ at a fixed $B_i = 0.01$ T, up to $30$ K in the temperature scale in the two scenarios. Note that cooling or heating can be predicted depending on the model considered.}
\label{mceminus100}
\end{figure}
The magnetocaloric effect (MCE) refers to the temperature variation of a material in response to the application and removal of an external magnetic field. This effect can be characterized by the temperature variation $\Delta T_B$ as a function of the system's temperature for different values of the maximum magnetic field $B_{\text{max}}$ and the rotation frequency $\Omega$. The analyzed graphs evaluated the thermal response in the two distinct scenarios, $m_G\neq m^*$ and $m_G\equiv m^*$. For all graphs, regardless of the value of $\Omega$, it is observed that an increase in $B_{\text{max}}$ results in a greater magnetocaloric effect. The temperature variation $\Delta T_B$ increases as $B_{\text{max}}$ increases. This occurs because a stronger magnetic field amplifies the system’s thermal response, inducing a greater redistribution of the electron energy.

Rotation significantly affects the behavior of the MCE, with distinct patterns for positive and negative values of $\Omega$. For positive values of $\Omega$, $\Delta T_B$ increases approximately linearly with temperature, indicating a heating effect, and is greater for $50$ GHz than $100$ GHz. For negative values of $\Omega$, $\Delta T_B$ takes on negative values at higher temperatures, indicating a cooling effect, with cooling being slightly stronger for $-50$ GHz compared to $-100$ GHz. This difference can be interpreted as a combined effect of rotation and the magnetic field, which alters the electron density of states and modifies their thermal response.

When comparing the sets of graphs for positive $\Omega$'s, it is noted that when $m_G\neq m^*$, the temperature variation is smaller than in the case where the masses are equal. This suggests that, when considering different masses, the interaction between electronic states and the magnetic field is reduced, limiting the magnetocaloric effect. On the other hand, when the masses are equal, the system’s thermal response to the magnetic field is more intense, resulting in greater variations in $\Delta T_B$. The cooling effect is not observed in this case, as we have seen that deviations in thermal properties are less prominent.

The analyses of magnetization $M(T,B)$ and the magnetocaloric effect $\Delta T_B$ highlight the intrinsic relationship between these phenomena. In particular, it is observed that for positive values of the rotation frequency $\Omega$, the decrease in magnetization is accompanied by an increase in $\Delta T_B$, characterizing a heating effect. This behavior remains consistent regardless of the adopted model, indicating that the interaction between the system's rotation and the external magnetic field directly influences the thermal response of the two-dimensional electron gas (2DEG).

On the other hand, for negative $\Omega$'s, the impact on magnetization and the magnetocaloric effect depends on the considered effective mass approach. When $m^* \equiv m_G$, the observed behavior is similar to the case of positive rotations, with the reduction in magnetization accompanying the system's temperature increase. However, when $m^* \neq m_G$, the magnetization response undergoes significant changes, directly impacting the manifestation of the magnetocaloric effect. In this configuration, cooling regimes emerge, where the temperature variation $\Delta T_B $ assumes negative values under specific conditions. This effect suggests that the difference between effective masses modifies the electronic density of states and how the system interacts with the magnetic field, influencing thermal transfer and the magnetocaloric response. Obviously, the graphs are valid for values that do not violate the third law of thermodynamics. 

The MCE for the degenerate Landau level case, with $\Omega \equiv 0$ Hz, is investigated adequately in Ref. \cite{e20080557}. The degeneracy of Landau levels fundamentally alters the behavior of the MCE: while in the non-degenerate case, the system heats up with magnetization, cooling occurs in the degenerate case.

These results emphasize the importance of considering rotational effects and effective mass properties in investigating the thermodynamics of electronic systems under magnetic fields, opening new possibilities for thermal modulation in semiconductor materials and quantum devices. Additionally, the magnetocaloric effect (MCE) can be used to probe the most suitable model for describing two-dimensional electron gases (2DEGs) under inertial effects. 

In the case of electrons in metals, there is no distinction between masses since $m^* = m_e$. Therefore, the behavior described for the case $m^* = m_G$ is the only relevant one, with all plots qualitatively similar to those studied here for GaAs, differing only quantitatively.

\FloatBarrier
\section{Concluding remarks}\label{sec5}

This work investigated the thermodynamic properties of a two-dimensional electron gas (2DEG) in a rotating medium under the Sagnac effect and a uniform magnetic field. The focus was on analyzing the impact of rotation and the distinction between effective mass ($m^*$) and gravitational mass ($m_G$) on energy levels and the system's thermodynamic responses.

The results show that rotation, even in the absence of a magnetic field, breaks the degeneracy of electronic states, and its presence significantly alters the Landau levels. Furthermore, introducing $m_G \neq m^*$ modifies the magnetization and favors cooling regimes in the magnetocaloric effect (MCE).

Rotation and the magnetic field strongly influence internal energy, specific heat, and Helmholtz free energy, with distinct effects depending on the $\Omega$ value. The system's entropy follows the Third Law of Thermodynamics, but its behavior changes with rotation due to the redistribution of electronic states. Magnetization strongly depends on the rotation frequency, with sign inversions in specific regimes when $m_G \neq m^*$. The magnetic field enhances the MCE and can lead to either heating or cooling depending on the rotation and the relation between $m_G$ and $m^*$.

These findings highlight the interaction between the Sagnac effect, the magnetic field, and the mass distinction in the thermodynamics of the 2DEG. Introducing $m_G$ directly impacts the system's responses, suggesting that its consideration is essential for more comprehensive models. Further investigation, including first-principles calculations and experiments, could provide deeper insights into this mass distinction and its implications for semiconductor systems and quantum devices.

\section*{Funding}

This work was partially supported
by the Brazilian agencies, CNPq, FAPEMIG, and FAPEMA: C.  Filgueiras and M. Rojas acknowledge FAPEMIG Grant No. APQ 02226/22. C. Filgueiras acknowledges  CNPq Grant No. 310723/2021-3, and M. Rojas acknowledges CNPq Grant No. 317324/2021-7. Edilberto O. Silva acknowledges the support from grants CNPq/306308/2022-3, FAPEMA/UNIVERSAL-06395/22, and FAPEMA/APP-12256/22. V. T. Pieve thanks for the master's scholarship provided by FAPEMIG.

\section*{Data Availability Statement}

No new data were created or analyzed in this study. Data sharing does not apply to this article.

\section*{Conflicts of Interest}

The authors declare no conflicts of interest.

\appendix

\section{Energy levels: $(n,l)\rightarrow(n_{+},n{-})$}\label{A}

We start by considering the following substitutions:
\[ \frac{\omega_{c}}{2} + \frac{m_{G}\Omega}{m^{*}} \equiv \frac{\varpi}{2} \quad \text{and} \quad \sqrt{\omega_{c}^{2} + 4m_{G}\Omega^{2} \frac{(m_{G} - m^{*})}{m^{*2}} + 4\omega_{c}\Omega \frac{(m_{G} - m^{*})}{m^{*}} } \equiv 2\Upsilon. \]
Taking into account the energy spectrum (\ref{Energyspectrum}), they yield
\begin{eqnarray}
 E = \hbar \Upsilon \left[ 2n + |\ell| + 1 \right] + \hbar \ell \frac{\varpi}{2}. \label{enl}
\end{eqnarray}
Making the change $ n_{+} = \frac{1}{2} \left( 2n + |\ell| + \ell \right) $ and $ n_{-} = \frac{1}{2} \left( 2n + |\ell| - \ell \right) $, with $ n_{\pm} = 0, 1, 2, \ldots $, we arrive at
\[ n_{-} - n_{+} = {2n} + {\frac{|\ell|}{2}} - \frac{\ell}{2} - {2n} -\frac{|\ell|}{2} - \frac{\ell}{2} = -\ell. \]
That is,
\[ \ell = n_{+} - n_{-}. \]
We also have
\[ 2n = 2n_{+} - |\ell| - \ell, \]

\[ 2n = 2n_{-} - |\ell| + \ell.\]
Adding these two relations, we get at
\[ 2n+|\ell| = n_{+} + n_{-}= 2n_{+}-\ell. \]
This way, we can rewrite (\ref{enl}) as
\begin{align}
 E &= \hbar \Upsilon \left[ 2n_{+}-\ell + 1 \right] + \hbar \ell \frac{\varpi}{2}\;\nonumber \\ &=\hbar (2\Upsilon) \left[ n_{+} + \frac{1}{2} \right] - \hbar \ell \left(\hbar \Upsilon-\frac{\varpi}{2}\right).   \label{enl2}
\end{align}
Considering the definitions of $\Upsilon$ and $\varpi$ above, we arrive at the energy levels shown in the text in Eq. (\ref{nmasimenos}). 

\bibliographystyle{apsrev4-2}
\bibliography{reference}
\end{document}